\documentclass[twocolumn,prl]{revtex4}
\usepackage{graphicx}
\usepackage{amsmath}
\usepackage{color}
\usepackage[]{todonotes}
\usepackage[noend]{algpseudocode}
\usepackage[nothing]{algorithm}
\algrenewcommand{\algorithmicrequire}{\textbf{Input:}}
\algrenewcommand{\algorithmicensure}{\textbf{Output:}}
\newcommand{\us}{\underline{~}}
\algrenewcommand{\textproc}{\textsf}

\usepackage{mathtools}
\DeclarePairedDelimiter\ceil{\lceil}{\rceil}
\DeclarePairedDelimiter\floor{\lfloor}{\rfloor}

\begin{document}
\title{Equivalence of several generalized percolation models on networks}
\author{Joel C. Miller}
\affiliation{Monash University, Melbourne, VIC Australia and Institute for Disease Modeling, Bellevue, WA USA}
\date{\today}
\begin{abstract}
In recent years, many variants of percolation have been used to study network structure and the behavior of processes spreading on networks.  These include bond percolation, site percolation, $k$-core percolation, bootstrap percolation, the generalized epidemic process, and the Watts Threshold Model (WTM).  We show that --- except for bond percolation --- each of these processes arises as a special case of the WTM and bond percolation arises from a small modification.  In fact  ``heterogeneous $k$-core percolation'', a corresponding ``heterogeneous bootstrap percolation'' model, and the generalized epidemic process are completely equivalent to one another and the WTM.  We further show that a natural generalization of the WTM in which individuals ``transmit'' or ``send a message'' to their neighbors with some probability less than $1$ can be reformulated in terms of the WTM, and so this apparent generalization is in fact not more general.  Finally, we show that in bond percolation, finding the set of nodes in the component containing a given node is equivalent to finding the set of nodes activated if that node is initially activated and the node thresholds are chosen from the appropriate distribution.  A consequence of these results is that mathematical techniques developed for the WTM apply to these other models as well, and techniques that were developed for some particular case may in fact apply much more generally.
%
\end{abstract}

\maketitle
%
%
\section{Introduction}
To understand processes spreading on a static network $G$, researchers frequently investigate how $G$ behaves under percolation.  Percolation comes in many flavors, and the information we gain depends on which variety we choose.  Most frequently, we study bond or site percolation, but researchers have also found that $k$-core percolation, bootstrap percolation, the generalized epidemic process, and the Watts threshold model (WTM) provide valuable insights~\cite{watts:WTM,goltsev:kcore,chalupa:bootstrap,adler:bootstrap,janssen:GEP,bizhani:GEP}.  These processes are closely related, and indeed similar mathematical approaches have been used to study several of these processes~\cite{miller:contagion,gleeson:cascades}.  Our main result is that all of these (and some related) processes can be derived as special cases of the WTM, and in fact several of these are completely equivalent to the WTM.

Much of the motivation for studying percolation processes comes from trying to understand spreading processes in networks.  If we consider systems in which nodes change status in response to the status of their neighbors, and the potential path of statuses they can have is acyclic (that is they can never return to a previous status), then many variants of percolation can be applied.  This is commonly used for SIR disease, in which an individual can be infected by an infected neighbor.  However, much recent work has focused on the spread of ``social contagion'' or ``complex contagions''~\cite{centola:weakness,centola:cascade} in which multiple transmissions may be required in order to cause ``infection''.  Sometimes this is presented as assigning each node a threshold $r_u$ such that $u$ becomes infected once $r_u$ neighbors are ``infected''.  Other times this is presented as a reduction (or increase) in the probability that a neighbor will transmit as an individual encounters more infected individuals.  This models the idea that after hearing seemingly independent ``confirmation'' of a rumor people may be more likely to believe and spread it, or after seeing multiple people engaging buying a product someone is more likely to perceive a consensus and buy the product as well.  Some experimental evidence of this has been found~\cite{centola:experiment,osullivan2015mathematical}.

We briefly review the processes we will study: In bond percolation, some edges are independently selected with uniform probability $p$ to be retained while the remaining edges are deleted (with probability $1-p$).  Similarly in site percolation, some nodes are  randomly selected with probability $p$ and the remaining nodes are deleted.  Typically our interest is in identifying the nodes in the connected components of the residual network, and whether a ``giant'' component exists (that is a component whose size is proportional to the network size in the infinite network limit).  
 
\emph{Bond percolation} and \emph{site percolation} often show up in the study of SIR disease spread where a single transmission suffices to cause infection~\cite{cardy:percolation,grassberger:percolation,kuulasmaa:bondperc,meester:bounds,moore2000epidemics,allard,sander:percolation,serrano:prl,meyers:contact,kenah:EPN,miller:random_clustered, miller:ebcm_overview}.  There is an exact equivalence between the spread of an SIR disese and bond percolation, and so much has been learned about the threshold, scaling properties, and dynamics of an SIR disease by studying the corresponding percolation odel.  This percolation equivalence is based on the fact that an edge either exists or does not in percolation, while in disease spread if the edge transmits, the receiving node becomes infected.

In \emph{$k$-core percolation} all nodes with degree less than some specified $k$ are removed.  This removal may reduce some nodes' degrees below $k$.  If so, these are removed.  This ``pruning'' process repeats until reaching a state in which all nodes have degree at least $k$.  This remaining network is called the ``$k$-core'' of the network.  It is seen to have hybrid phase transitions, with a square root-type scaling on one side of a transition followed by a discontinuous jump~\cite{goltsev:kcore,baxter:dynamics}.  In a variant, ``heterogeneous $k$-core'' percolation~\cite{baxter:heterogeneous_kcore}, each node is assigned its own threshold value and deleted if its degree goes below the threshold.   We note that many authors have used the term ``bootstrap percolation'' to denote $k$-core percolation, and indeed this appears to be the original term~\cite{goltsev:kcore,chalupa:bootstrap,adler:bootstrap}, but we reserve ``bootstrap percolation'' for a closely related dual process.  The $k$-core has been applied to many problems, including understanding failure of a physical system under strain~\cite{shi2005strain}, network visualization~\cite{alvarez2005large}, identification of the component of a network responsible for establishing a disease~\cite{kitsak2010identification}, and more generally for understanding the structure of a network~\cite{dorogovtsev2006k}.

In \emph{bootstrap percolation} (introduced in~\cite{adler:bootstrap} where it is called ``diffusion percolation'') a collection of nodes is initially ``activated''.  Then any inactive node with at least $m$ active neighbors becomes active.  The process repeats until all remaining inactive nodes have fewer than $m$ active neighbors.  It was initially introduced to model the spread of a water-filled crack in a rock.  It has received considerable study on lattices~\cite{schonmann1992behavior,cerf1999finite}, and its behavior in large random networks has been the subject of some more recent analysis~\cite{baxter2010bootstrap}.  Like $k$-core percolation it is seen to have a hybrid phase transition.
We introduce a natural generalization analogous to heterogeneous $k$-core percolation in which each node is assigned its own threshold.  This ``heterogeneous bootstrap percolation'' does not appear to have been studied previously.

In the \emph{generalized epidemic process} (GEP)~\cite{janssen:GEP,bizhani:GEP,chung2016universality}, we think of an infection spreading through the network.  If a node has a single infected neighbor, its probability of becoming infected is $p_1$.  If it escapes infection, but a second neighbor becomes infected, then its probability of becoming infected is $p_2$.  This repeats and the probability of successful transmission on the $m$-th neighbor's infection is $p_m$.  If $p_m=p$ for all $m$, then this is the network version of the classical Reed-Frost model~\cite{abbey:reedfrost} for a susceptible-infected-recovered disease~\cite{andersonMay}.  If $p_m$ decreases as $m$ increases, this could model decreasing susceptibility due to an improved immune response as exposures accumulate, or it could simply represent pre-existing heterogeneities in susceptibility that are revealed as the number of exposures increases.   An increasing $p_m$ would model some synergistic or cumulative effect of exposures as seen in ``complex contagions''~\cite{centola:cascade}.  For comparison with other models, we allow $p_m$ to depend on $d_u$, the degree of node $u$.

In the \emph{WTM}~\cite{watts:WTM,dodds:contagionPRL}, each node $u$ is assigned an individual threshold $r_u$ which we assume is assigned to $u$ independently at random, with a probability that may depend on its degree $d_u$.  The probability node $u$ has a given $r$ is given by $P(r_u=r|d_u) = q(r|d_u)$.  A node begins either active or inactive.  If an inactive node $u$ has at least $r_u$ active neighbors then it becomes active.  We assume that the initially active nodes may be chosen independently at random (which can be modelled by having $q(r|d)>0$ for some $r \leq 0$), or they may be chosen by some other rule, in which case we treat the set of initially active nodes as an input to the algorithm.  Often a common threshold $r^*$ is chosen so $P(r_u=r^*)=1$ or a common fraction $\rho^*$ is chosen so $P(r_u=\ceil{\rho^* d_u}|d_u) = 1$.  As described above, this is frequently used to model social contagions.  In~\cite{watts:WTM} it was conjectured that for a global cascade to occur from an infinitesimally small initial proportion active, a giant component of nodes with $r=1$ would need to exist.  This is true in random Configuration Model networks, but false in random clustered networks~\cite{miller:contagion}.  As with bootstrap and $k$-core percolation, this is known to exhibit hybrid bifurcations~\cite{miller:contagion}.

In these generalized percolation processes, typically we are interested in the final set of active nodes, but sometimes we may be interested in the temporal dynamics as these nodes become active~\cite{miller:contagion,baxter:dynamics}.  If we are interested in the temporal dynamics, then we must assign additional rules for how long it takes for a node to become active.  Although the timing will depend on the details of the additional rules, the final set of active nodes is uniquely determined once the network, thresholds, and initially active nodes are chosen.  For our purposes we focus just on the final state.

We will show that by appropriately choosing the distribution of $r$ and the initial set of active nodes, we can recover other versions of percolation from the WTM, including site percolation, $k$-core percolation, bootstrap percolation and the GEP.  Going a step further, we show that the heterogeneous $k$-core of a network, the deleted nodes in heterogeneous bootstrap percolation, and the set of ``infected nodes'' in the GEP are in fact all equivalent to the set of active nodes emerging from the WTM.  That is, given one model and the corresponding distribution of thresholds, we can define the distribution of thresholds of the other models to yield the same sets of nodes with the same probabilities.  A natural generalization of the WTM to consider has each node ``transmitting'' or ``passing a message'' with some fixed probability $T$.  We show that by modifying the threshold distribution the original WTM (with $T=1$) can recover the same outcomes as for any other $T_0<1$, and thus allowing for $T<1$ does not enlarge the set of possible outcomes.  

Finally, we investigate the relation with bond percolation.  If our interest in bond percolation is to identify the connected component containing a given node $u$, then we can find this component using the WTM with $u$ as the initially active node and appropriate threshold distribution.  To find all connected components, we can start the WTM with one initially active node, run it to completion, and then choose a remaining inactive node and rerun the WTM, iterating until no inactive nodes remain.  The set of nodes that are activated in each pass correspond exactly to the components found in bond percolation.

\section{Analysis}
We begin by explicitly describing an algorithm which implements the WTM.  Each node is assigned a weight $w$ uniformly between $0$ and $1$.  We use a model-dependent function \textsf{dist\us{}func} to convert the weight into a threshold $r$, possibly depending on the degree of the node.  Typically we choose the function to return the largest value $r_u$ such that $\sum_{r=-\infty}^{r_u-1} q(r|d) < w_u$.
If there are specified initially active nodes, they are given a threshold of zero.  Alternately we can allow the randomly assigned threshold to permit values $r_u\leq 0$ in which case these nodes are initially active, and the iterative process begins.  For each active node, we reduce the threshold of any inactive neighbor by $1$.  If a node's threshold reaches $0$ it activates.  Pseudocode for the algorithm is given in the appendix in figure~\ref{fig:WTM_alg}

Once the random thresholds and index nodes are set, the final outcome of the WTM is deterministic.  To show that the other percolation processes give the same behavior, we will show how to structure these processes to start from the same random weights $w_u$ and deterministically yield a final state that is identical to the state found by the WTM for some threshold distribution.

\subsection{Site Percolation}
\begin{figure}
1\framebox{\includegraphics[width=0.7\columnwidth]{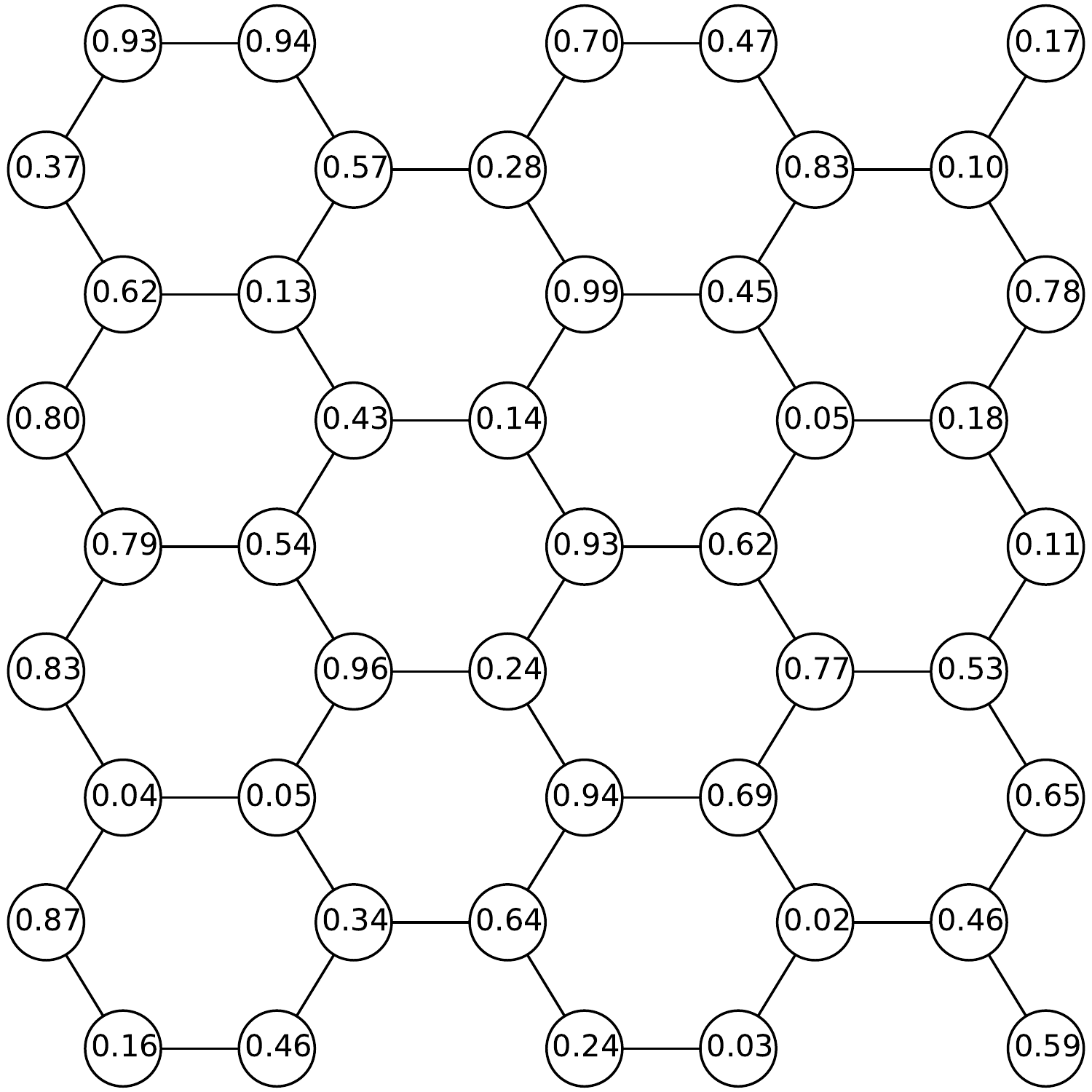}}
\\
\framebox{\includegraphics[width=0.45\columnwidth]{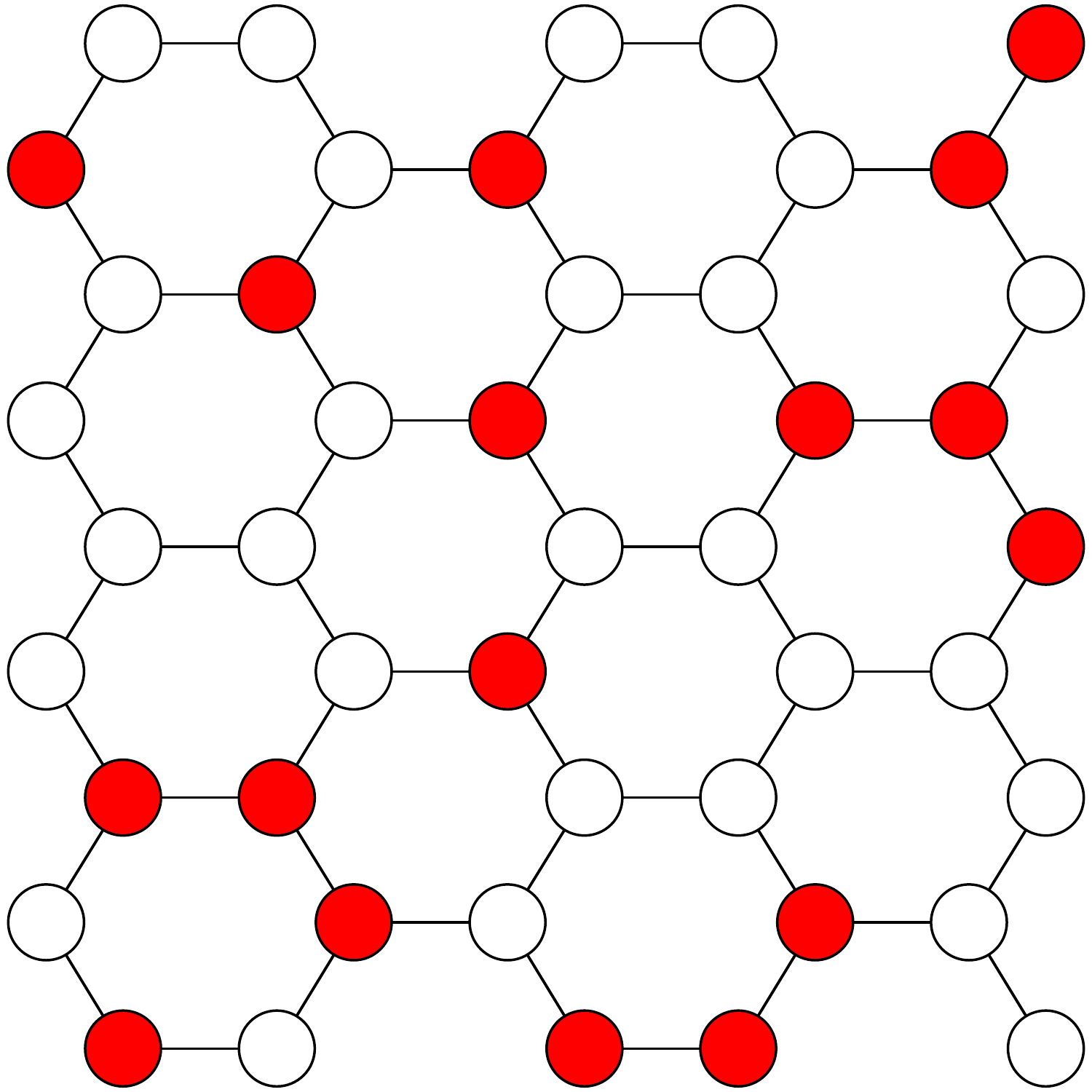}}
\hfill \framebox{\includegraphics[width=0.45\columnwidth]{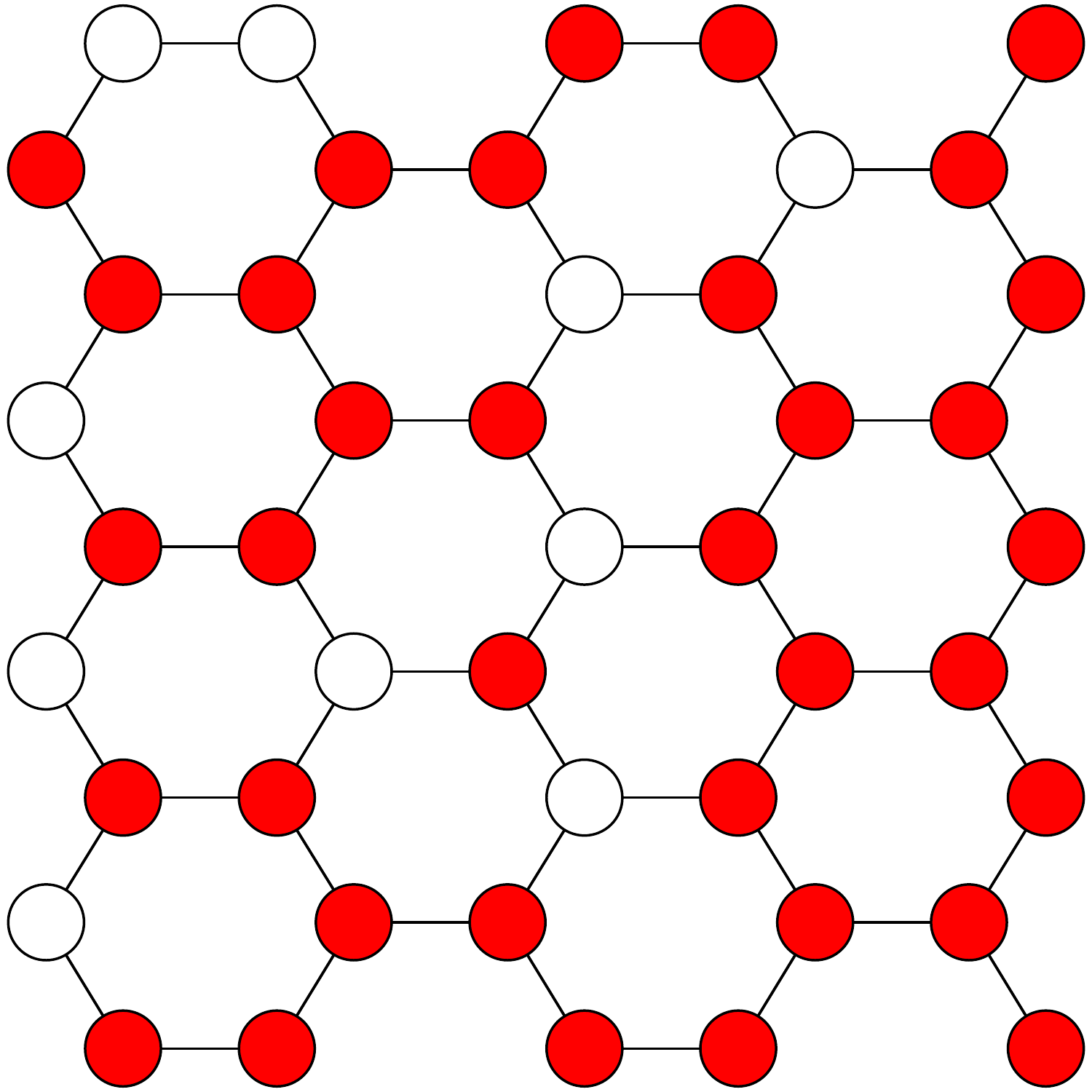}}
\\
\framebox{\includegraphics[width=0.45\columnwidth]{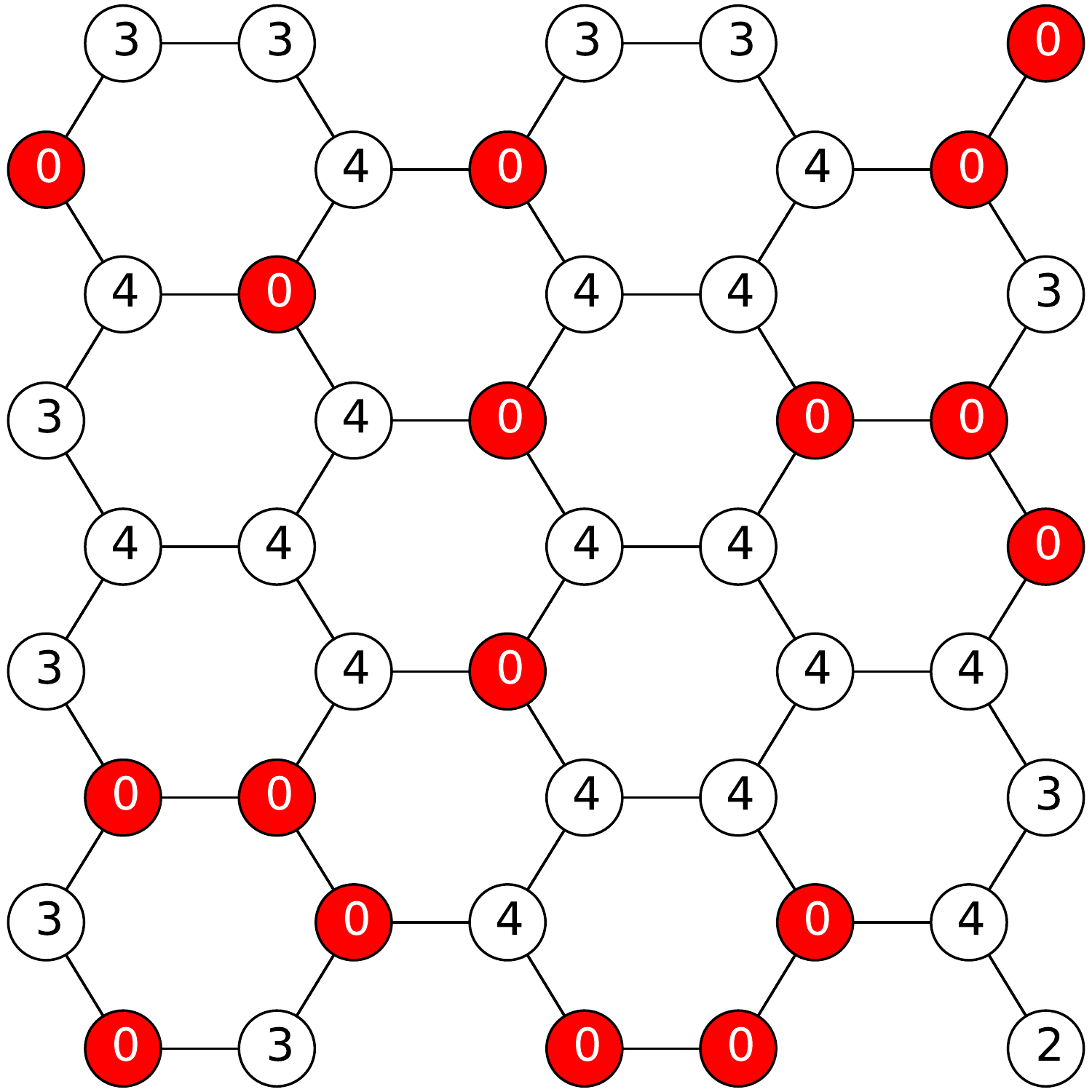}}
\hfill \framebox{\includegraphics[width=0.45\columnwidth]{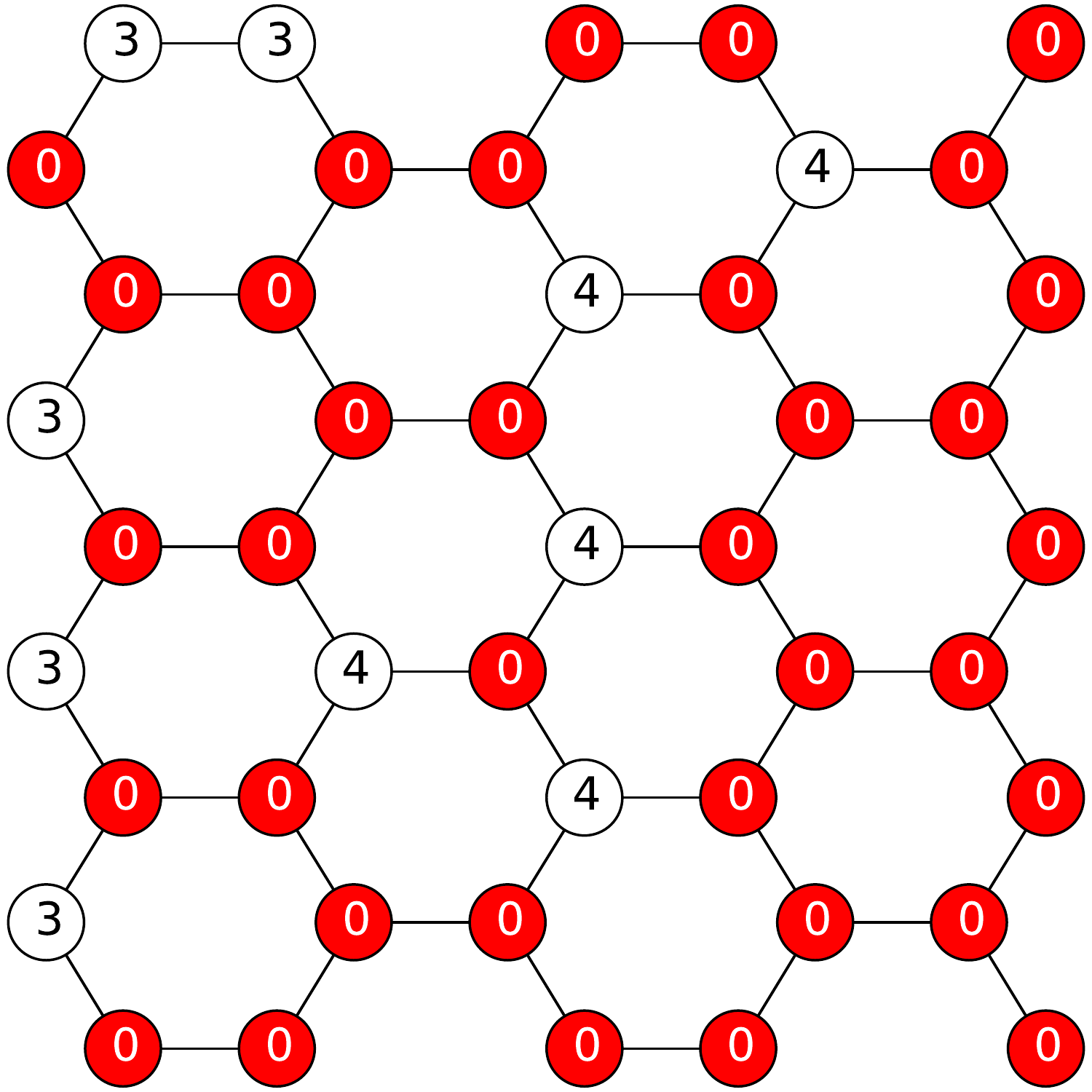}}
\caption{Comparison of site percolation and WTM on honeycomb lattice.
  Results for $p=0.4$ on left and $p=0.8$ on right.  (top) Each node
  is assigned a weight.  (middle) Site percolation: If the weight is less than $p$, the node is kept, otherwise it is deleted.
  (bottom) WTM: if the weight is less than $p$ it is given a threshold of $0$.  Otherwise it is given
  $d+1$.  Those with threshold $0$ are shown in color, and activate immediately.  Those with threshold larger than their
  degree are uncolored and never activate. }
\label{fig:site_vs_wtm}
\end{figure}

In site percolation, each node is retained with probability $p$ or deleted with probability $1-p$.  To simulate site percolation, we can generate a random number $w_u \in (0,1)$ independently and uniformly at random for each node $u$.  If $w_u<p$ (which occurs with probability $p$) we keep $u$, otherwise we delete it.  It is straightforward to see that this is identical to the algorithm in Figure~\ref{fig:WTM_alg} if the threshold is set to be $r_u=0$ whenever $w_u<p$ and $r_u=d+1$ otherwise.  In this case, with probability $p$ the node has threshold $0$ and so is initially active, while with probability $1-p$ it has threshold $d+1$, and so can never become active as it will have at most $d$ active neighbors.  Thus, nodes are retained in site percolation iff they are active in the WTM.  This is demonstrated in figure~\ref{fig:site_vs_wtm}. 

\subsection{$k$-core Percolation}

We now consider $k$-core and heterogeneous $k$-core percolation.  The classical $k$-core percolation is deterministic: each node with fewer than $k$ neighbors is deleted.  This iterates until all remaining nodes have at least $k$ neighbors among the remaining nodes.  To reproduce this with the WTM, we set $r_u= d_u-k+1$ regardless of $w_u$.

\begin{figure}
\includegraphics[width=0.48\columnwidth]{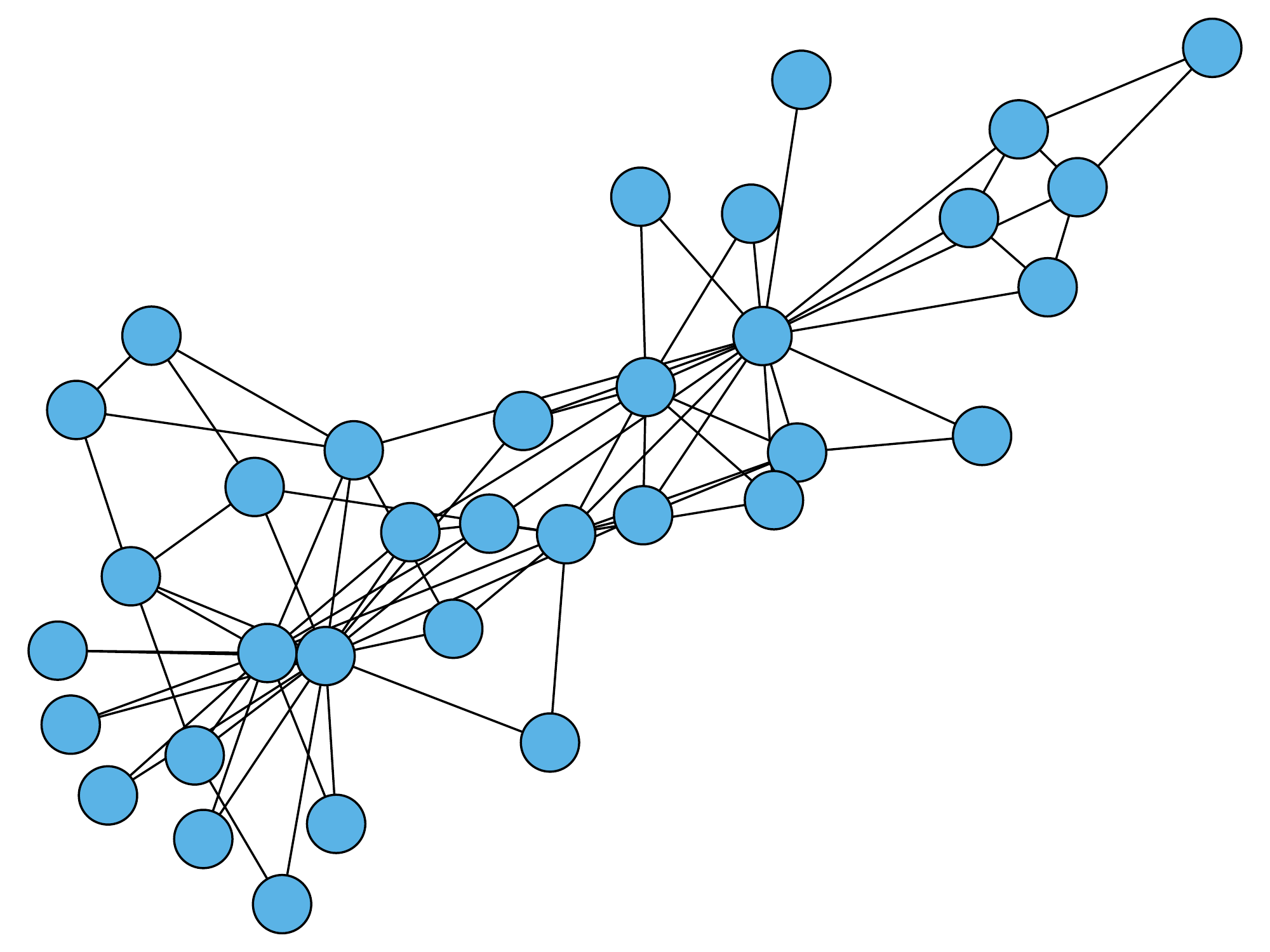}\hfill
\includegraphics[width=0.48\columnwidth]{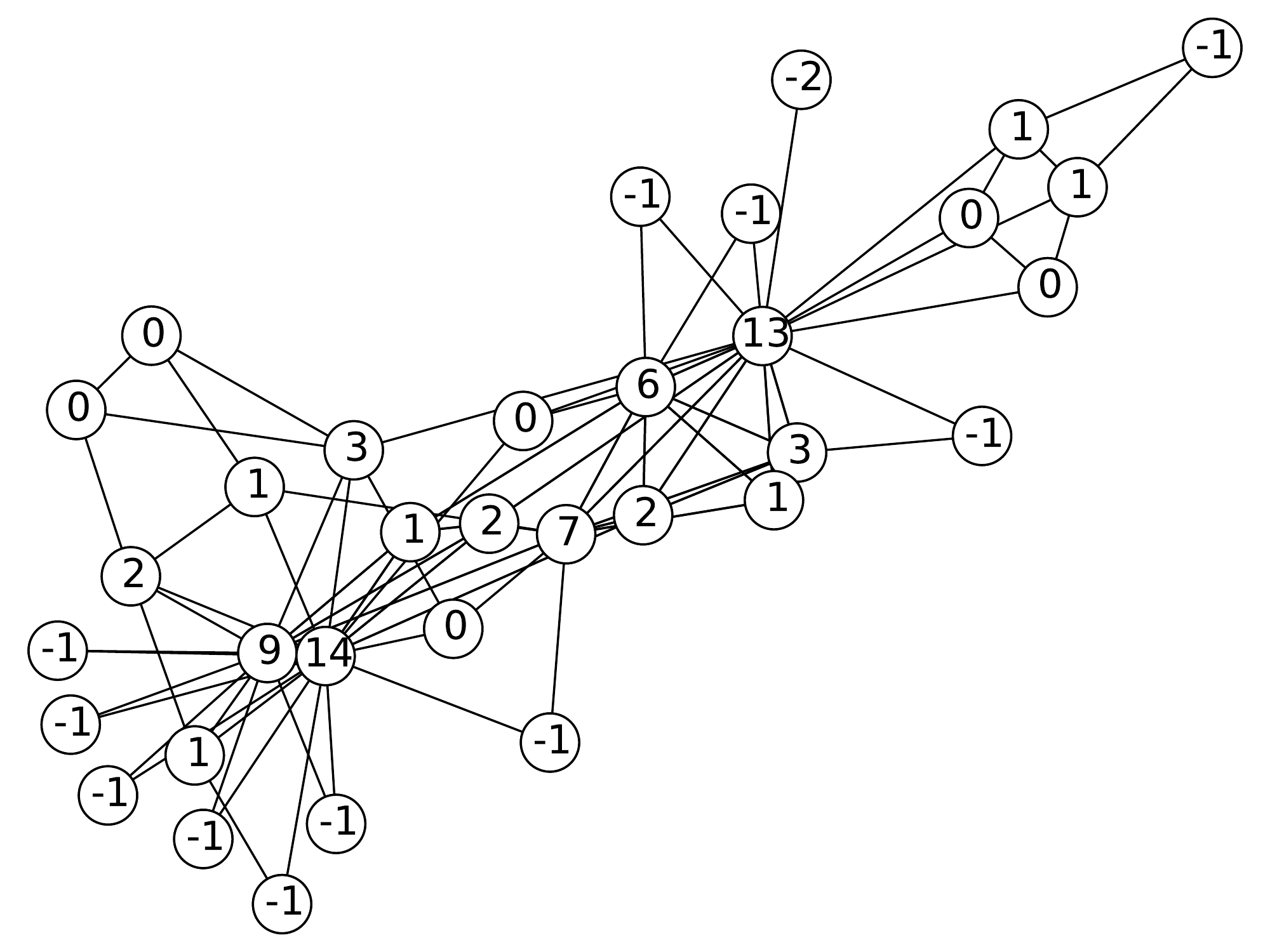}\\
\includegraphics[width=0.48\columnwidth]{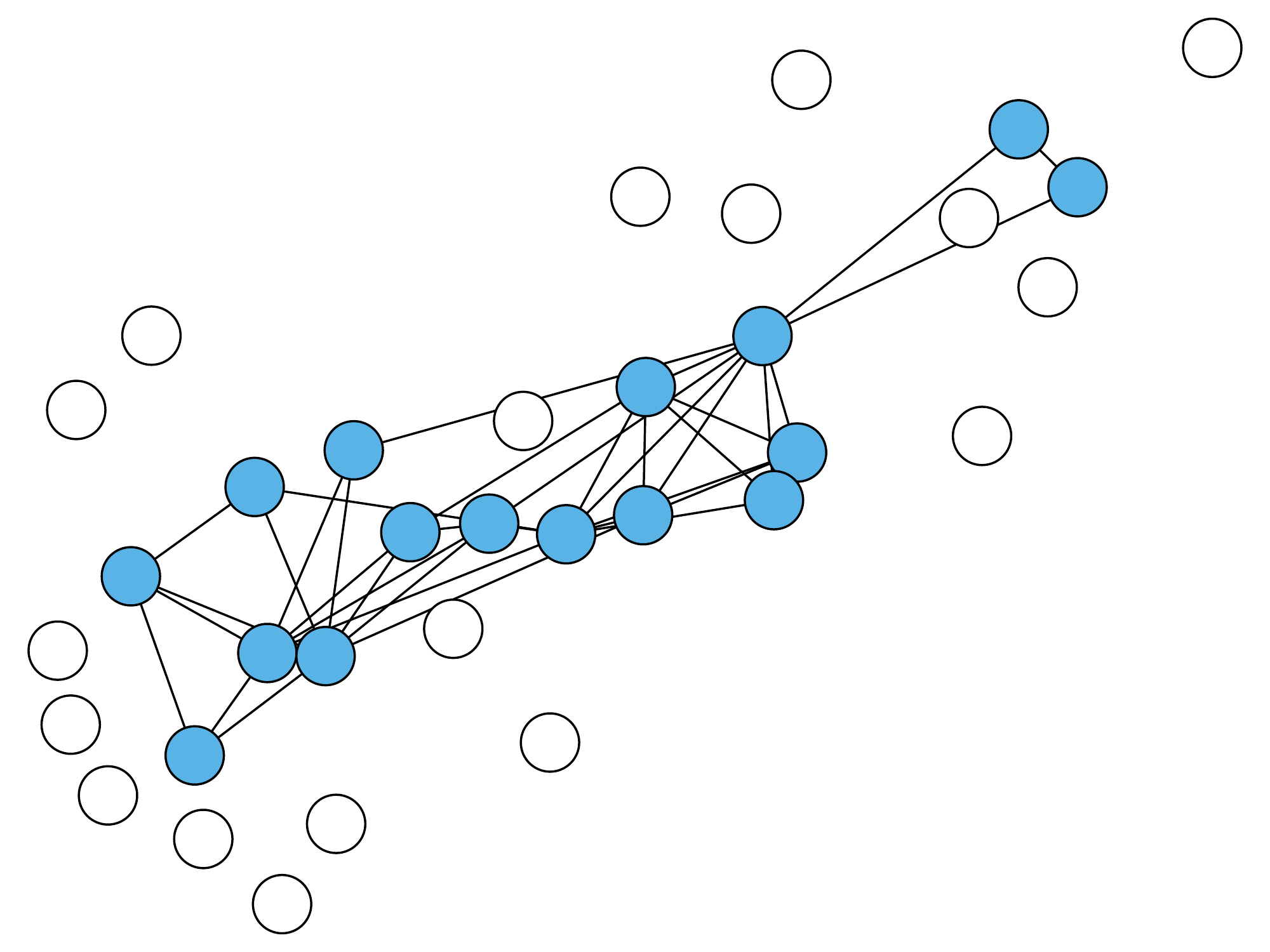}\hfill
\includegraphics[width=0.48\columnwidth]{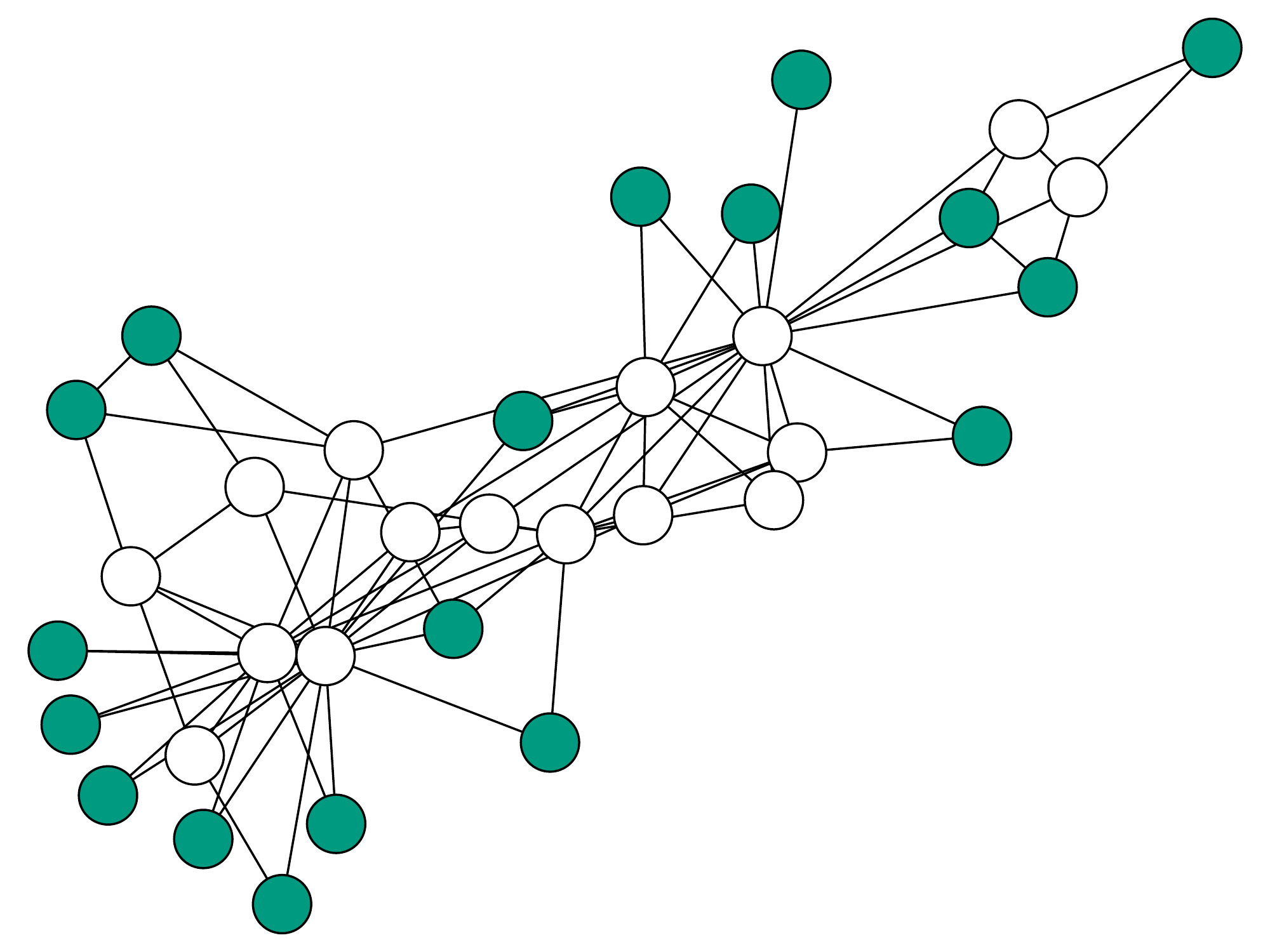}\\
\includegraphics[width=0.48\columnwidth]{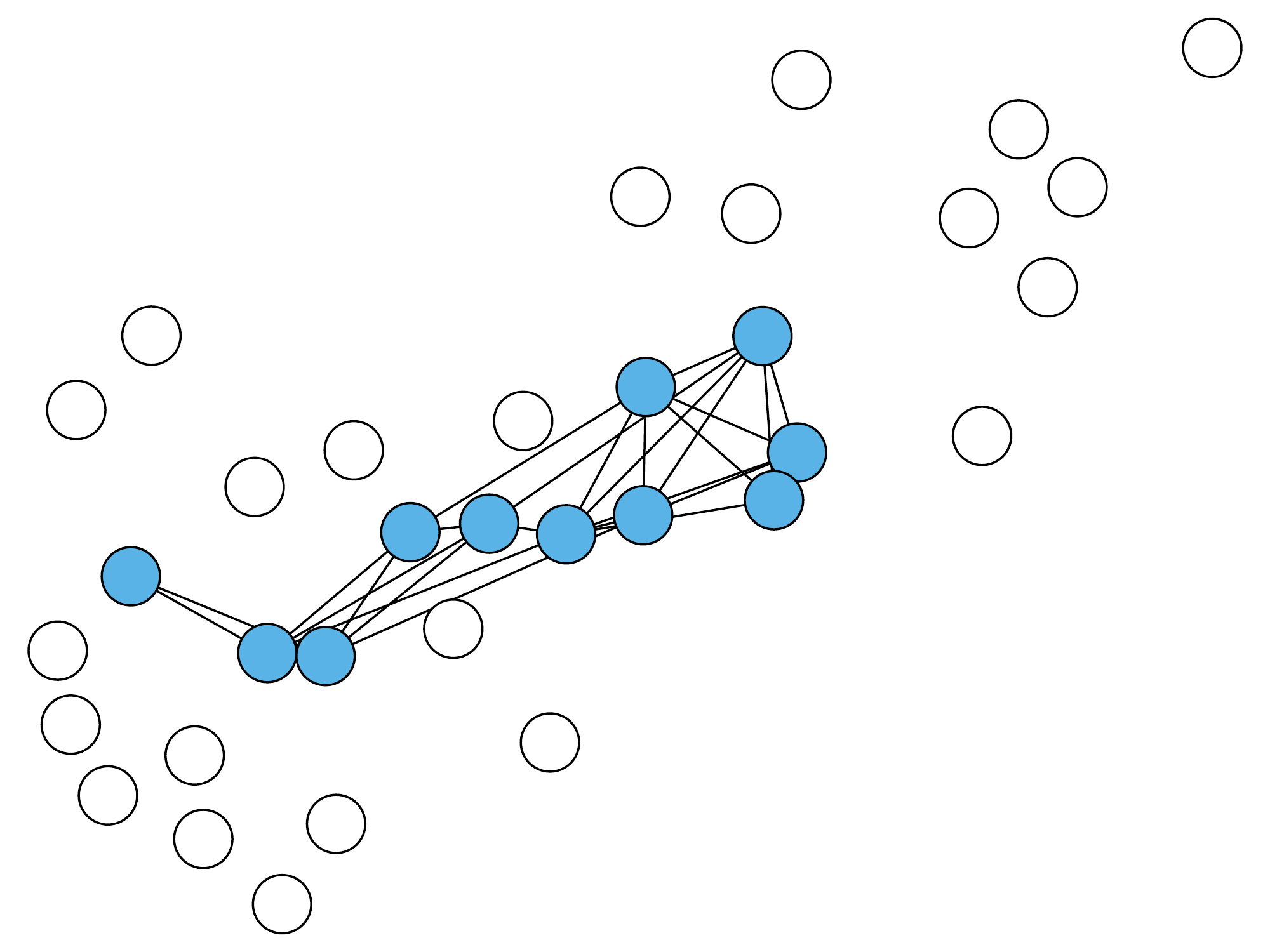}\hfill
\includegraphics[width=0.48\columnwidth]{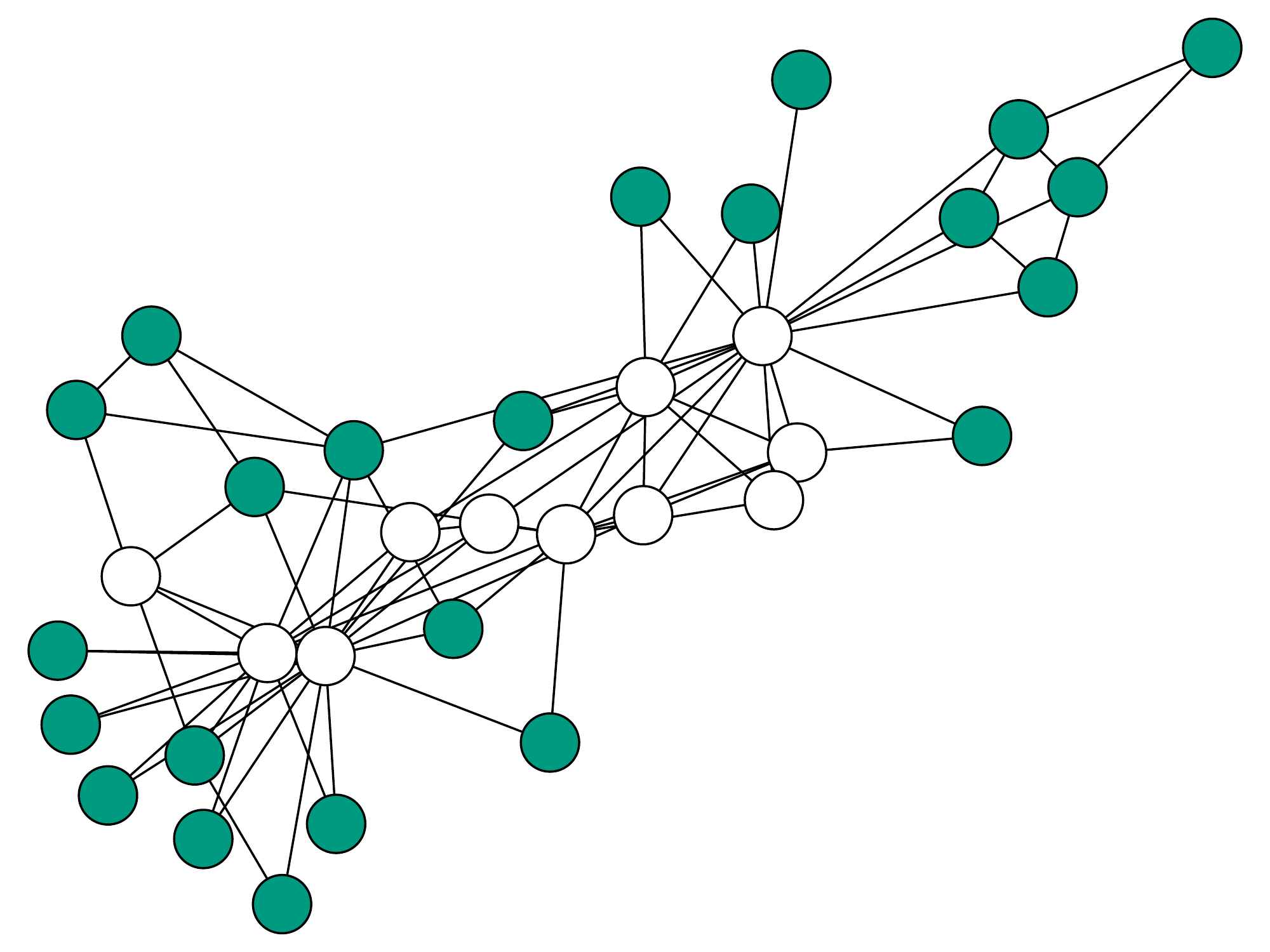}
\caption{Comparison of the first steps of $k$-core percolation and the WTM for the   karate club graph~\cite{zachary:karate}. (Left) $k$-core percolation with $k=4$.  (top) The original network.  (middle) The first step of $k$-core percolation.  (bottom) The second step.  (Right) The WTM (top) Thresholds of $d_u-4+1$.  (middle) Nodes with  thresholds $r_u\leq 0$ are activated. (bottom) The second step.  At each step, the activated nodes of the WTM are exactly the deleted nodes in $k$-core percolation.}
\label{fig:karateKcore}
\end{figure}

With this threshold, all nodes with $d_u<k$ activate  immediately in the WTM.  In $k$-core percolation these same nodes are immediately deleted.  For a given node $u$ not in this set, let the number of neighbors activated/deleted be denoted $n_u$.  In the WTM, any remaining node with $d_u-k < n_u$ then activates.  In $k$-core percolation, any node with $d_u-n_u<k$ is deleted.  Again, these nodes are the same.  Iterating as shown in figure~\ref{fig:karateKcore}, the set of activated nodes in the WTM is the set of deleted nodes in $k$-core percolation.

We can repeat this for heterogeneous $k$-core percolation.  We assign weights $w_u$ to each node and map that to a heterogeneous $k$-core threshold $k_u$.
We can map this weight to a WTM threshold such that if the node is assigned a given $k_u$, it is assigned $r_u = d_u-k_u+1$ for the WTM.  Then the WTM and heterogeneous $k$-core percolation are equivalent: a node is deleted in heterogeneous $k$-core percolation iff it is activated in the WTM.

\subsection{Bootstrap Percolation}
In bootstrap percolation, some initial nodes are activated, and nodes become active once they have at least $k$ active neighbors ($k$ is the same for all nodes).  This is similar to $k$-core percolation, but $k$-core percolation is subtractive while bootstrap percolation is additive~\cite{baxter:heterogeneous_kcore,baxter2010bootstrap}.

We consider bootstrap percolation with a set $I_0$ of initially active nodes, and compare it to the WTM with $r_u=k$ for all nodes except the nodes in $I_0$ which are initially active.  Following a similar argument to the WTM/$k$-core percolation equivalence, we see that with this definition, the WTM adds nodes to the system exactly when bootstrap percolation does.

If we consider heterogeneous bootstrap percolation, then a similar argument also shows that it is equivalent to the WTM.  Because of the correspondence between the WTM and heterogeneous $k$-core percolation, this means that heterogeneous bootstrap percolation is equivalent to heterogeneous $k$-core percolation, with the deleted nodes in heterogeneous $k$-core percolation matching the activated nodes in bootstrap percolation.  

At first glance, this contrasts with observations of~\cite{baxter:heterogeneous_kcore}.  They showed that the $k$-core and the activated nodes in bootstrap percolation are not the same and can have different internal structure.  In fact, the distinction between the two turns out to be that the nodes defined to be active for the bootstrap version are the nodes deleted in the $k$-core version.  They are complementary processes.  Any behavior observed in heterogeneous $k$-core percolation can be observed in heterogeneous bootstrap percolation the inactivated nodes of heterogeneous bootstrap percolation, while any behavior observed in the activated nodes of bootstrap percolation can be found in the deleted nodes of $k$-core percolation.  This equivalence is previously known~\cite{adler:bootstrap}.

\begin{figure}
\includegraphics[width=0.5\columnwidth]{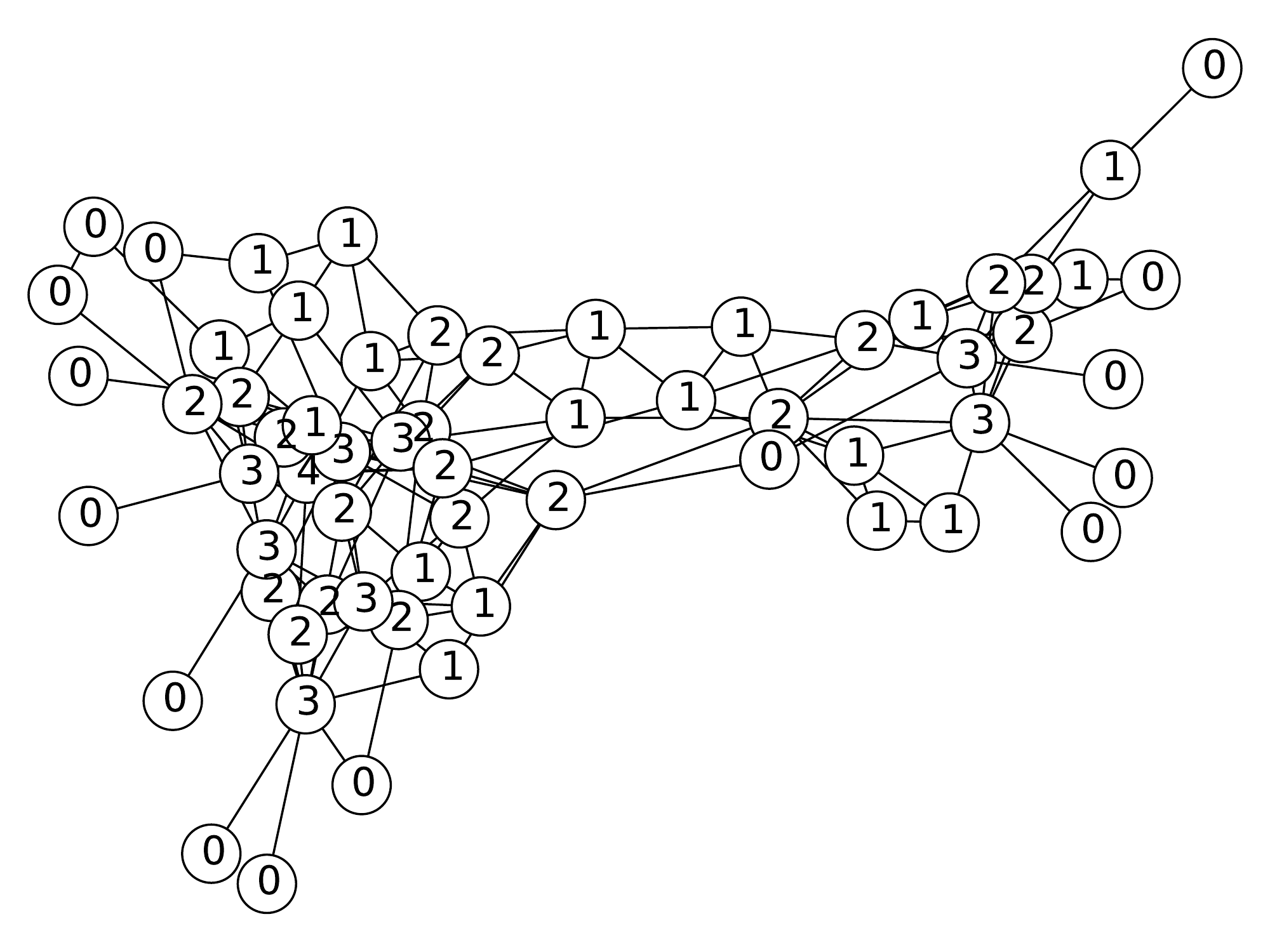}%
\includegraphics[width=0.5\columnwidth]{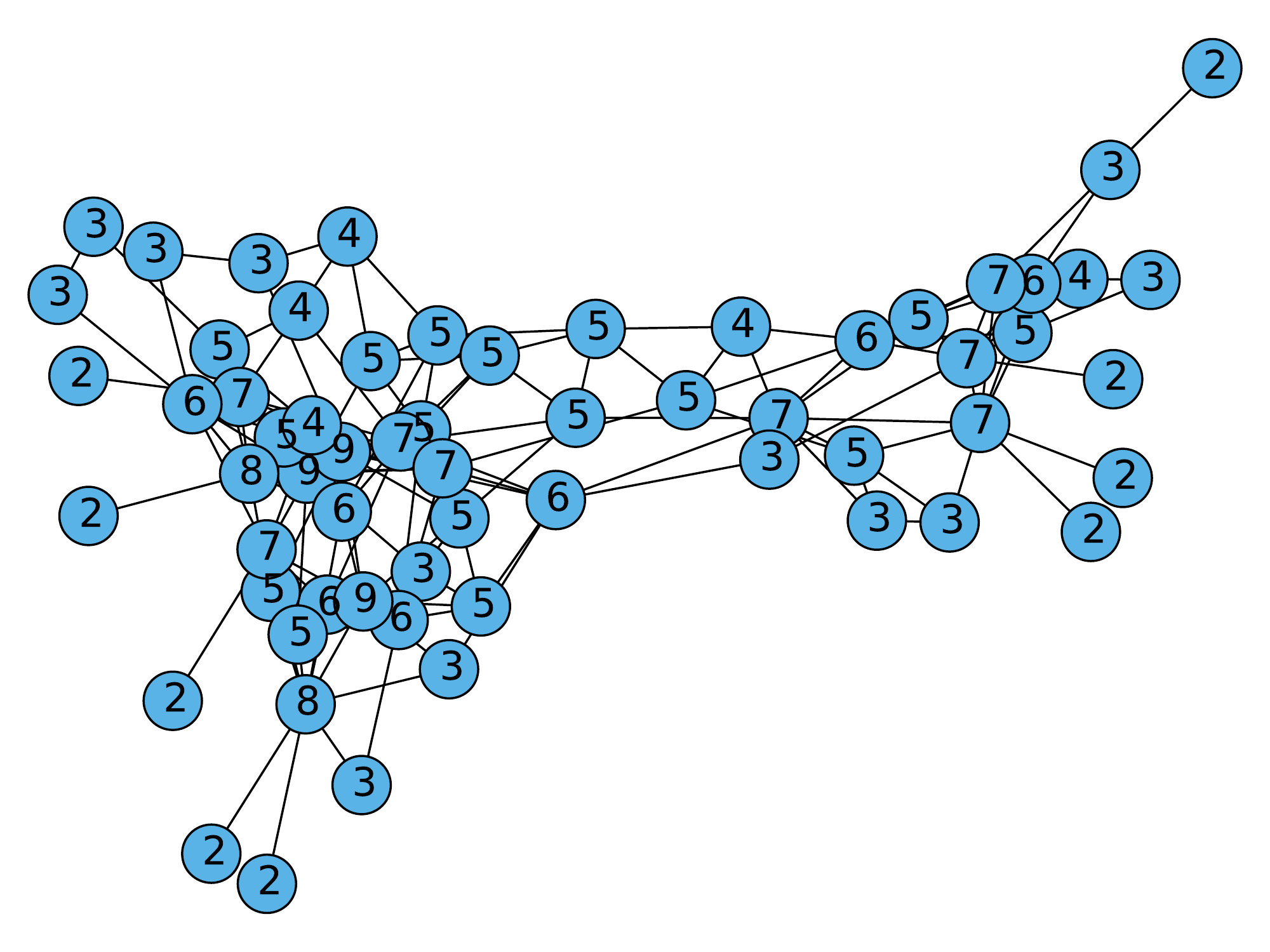}\\%
\includegraphics[width=0.5\columnwidth]{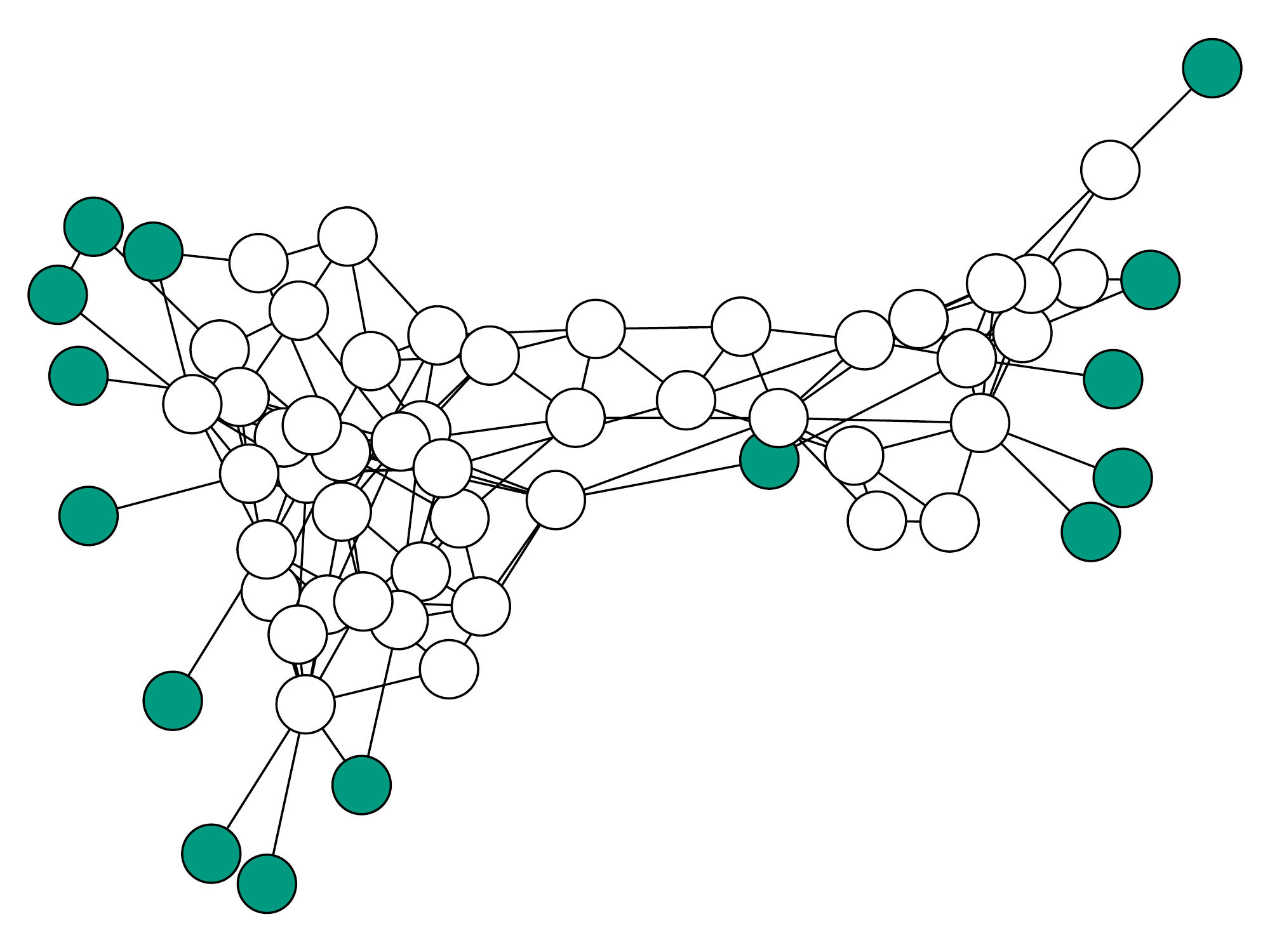}%
\includegraphics[width=0.5\columnwidth]{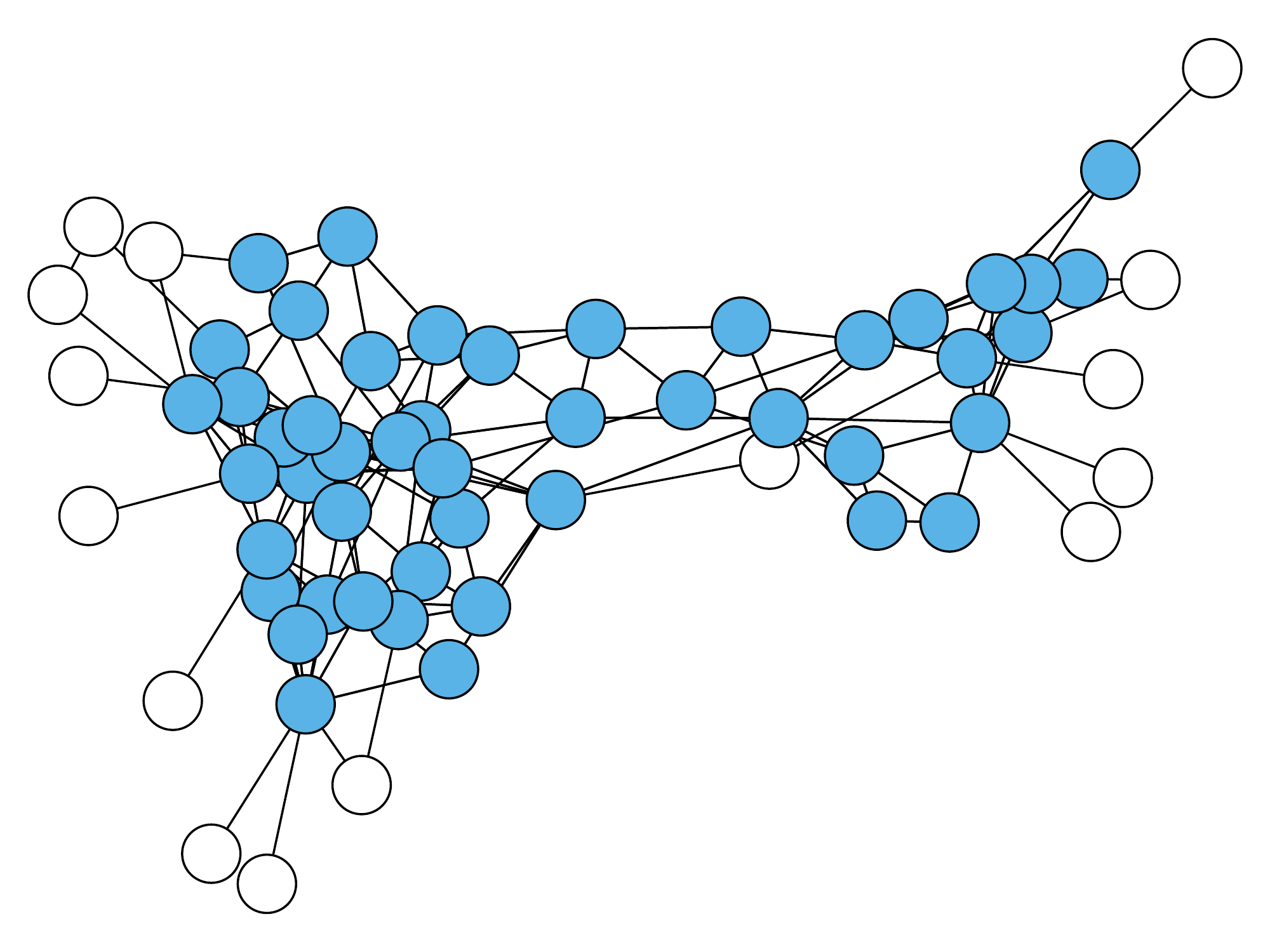}\\%
\includegraphics[width=0.5\columnwidth]{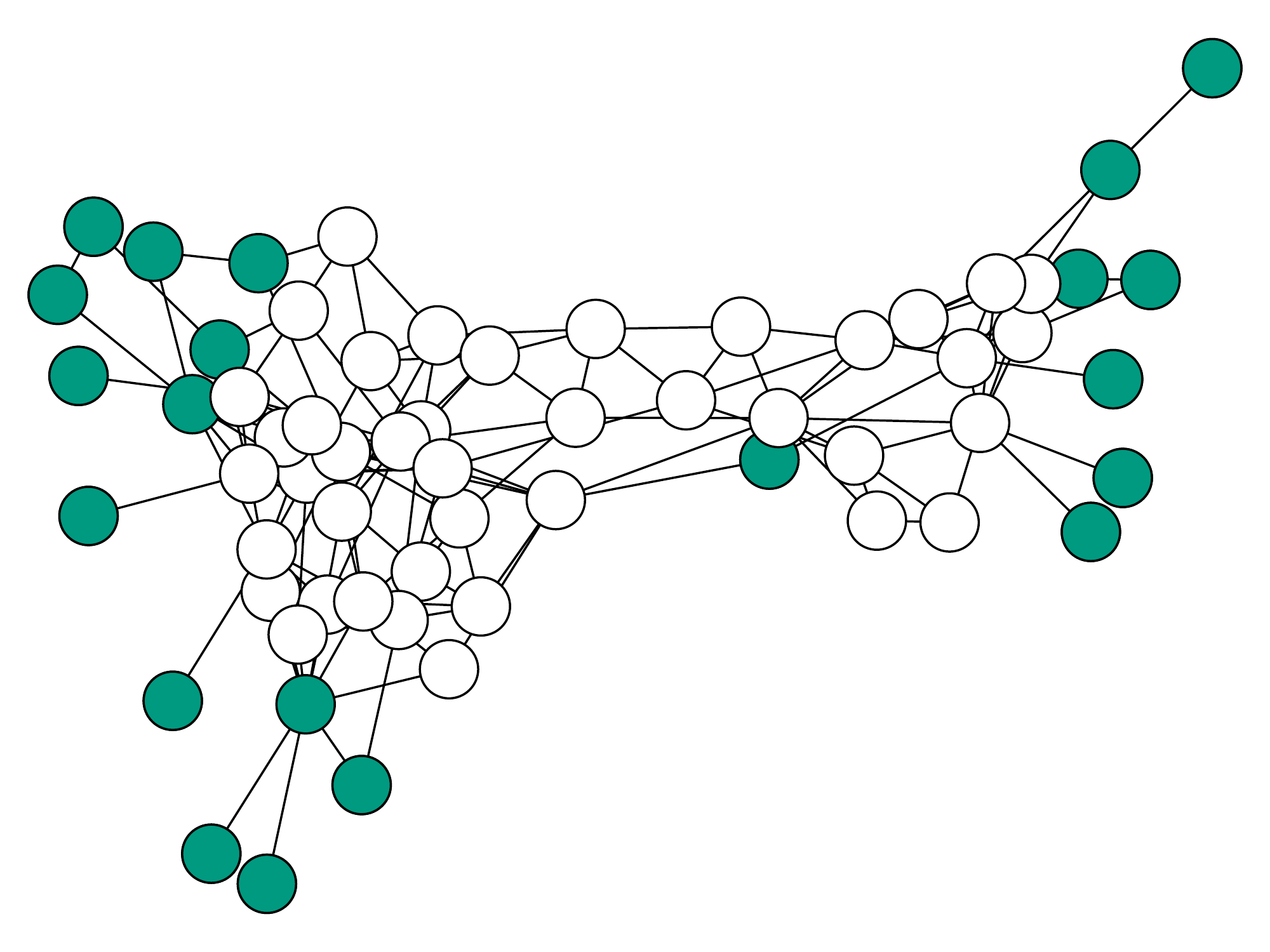}%
\includegraphics[width=0.5\columnwidth]{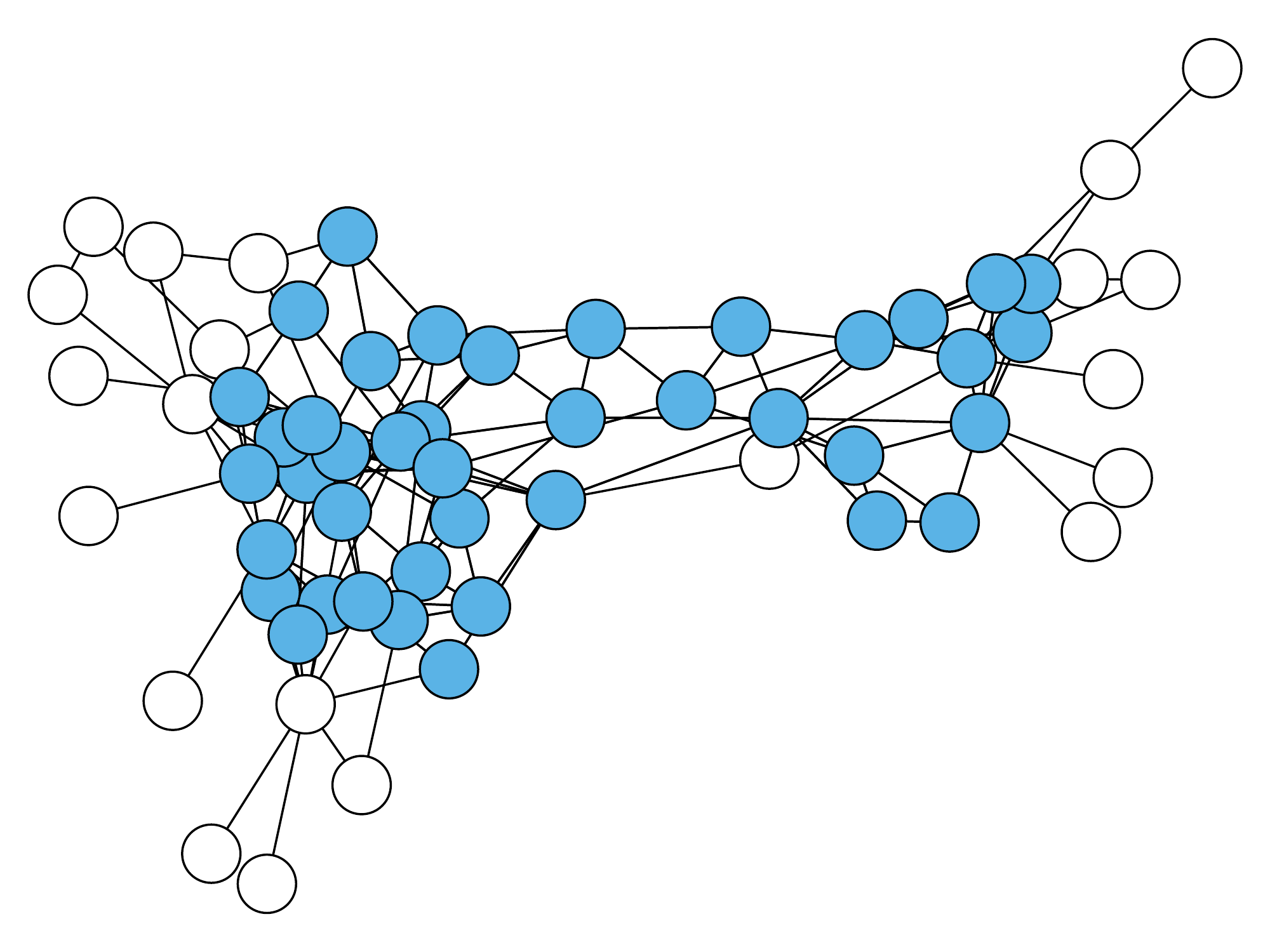}%
\caption{A comparison of heterogeneous bootstrap and heterogeneous
  $k$-core percolation for the social network of dolphins observed
  by~\cite{lusseau:dolphins}. (Left) Heterogeneous bootstrap percolation: (top) thresholds for activation, $\floor{d_u/3}$.  (middle) first step: all nodes of degree $1$ or $2$ are activated.  (bottom) second step: nodes that now reach their threshold are activated. (Right) Heterogeneous $k$-core percolation: (top) thresholds for deletion, $d_u-\floor{d_u/3}+1$.  (middle) first step: all nodes of degree $1$ or $2$ are deleted.  (bottom) second step: nodes that now reach their threshold are deleted.  The nodes deleted at each stage of $k$-core percolation correspond exactly to the nodes activated at the same stage of bootstrap percolation.}
\label{fig:dolphins}  
\end{figure}

Figure~\ref{fig:dolphins} demonstrates the equivalence between heterogeneous bootstrap and heterogeneous $k$-core percolation.

\subsection{Generalized Epidemic Process}
We now consider the generalized epidemic process (GEP)~\cite{janssen:GEP,bizhani:GEP} for which the $m$-th ``infected'' neighbor infects node $u$ (given that the previous $m-1$ did not) with probability $p_m(d_u)$.  Our approach resembles the ``Sellke construction''~\cite{sellke1983asymptotic} of a simple epidemic model in a fully-mixed population.  In a standard fully-mixed epidemic simulation, an individual that is susceptible at the start of a short time interval becomes infected with a probability proportional to the number of infected individuals.  In the Sellke construction formulation, however, we assume we know in advance for each individual the cumulative amount of exposure it will receive before becoming infected (this is a random number chosen from an exponential distribution).  We then begin the spread with some initial infections and when (or if) the exposure reaches that threshold the individual becomes infected.

We will now study the network-based GEP by a similar approach.
The probability the first $m-1$ infected neighbors do not infect $u$ but the $m$-th does is $p_m(d_u) \prod_{\hat{m}=1}^{m-1} (1-p_{\hat{m}}(d_u))$.  We simply assign a random number $w_u \in (0,1)$ and map this to $m_u$.  Thus for any given node, it will become infected upon the infection of its $m$-th neighbor with probability $p_m(d_u) \prod_{\hat{m}=1}^{m-1} (1-p_{\hat{m}}(d_u))$ independently of other nodes and independently of whether we will calculate $m_u$ in advance or simply accept or reject infection with probability $p_m(d_u)$ as it accumulates infected neighbors.

For the WTM we use the same mapping from $w_u$ to $r_u$ so $r_u=m_u$.  The node $u$ activates exactly after the $r_u$-th neighbor activates, while in the GEP $u$ is infected at exactly the same step.  Thus any GEP can be expressed as a WTM.  Showing the inverse is straightforward, and so the GEP and WTM are equivalent.  If we do not allow $p_m$ to depend on $d_u$ (as in the original version), then this is a special case of the WTM.

\subsection{Bond Percolation}
We finally consider Bond Percolation.  Typically in bond percolation, we can consider the edges in any order, choosing to keep each edge with probability $p$ or delete it with probability $1-p$ independently of the others.  We then identify the connected components of the network.

\begin{figure}
\begin{center}
~\mbox{}\hfill\raisebox{-0.49\height}{\mbox{\includegraphics[width=0.46\columnwidth]{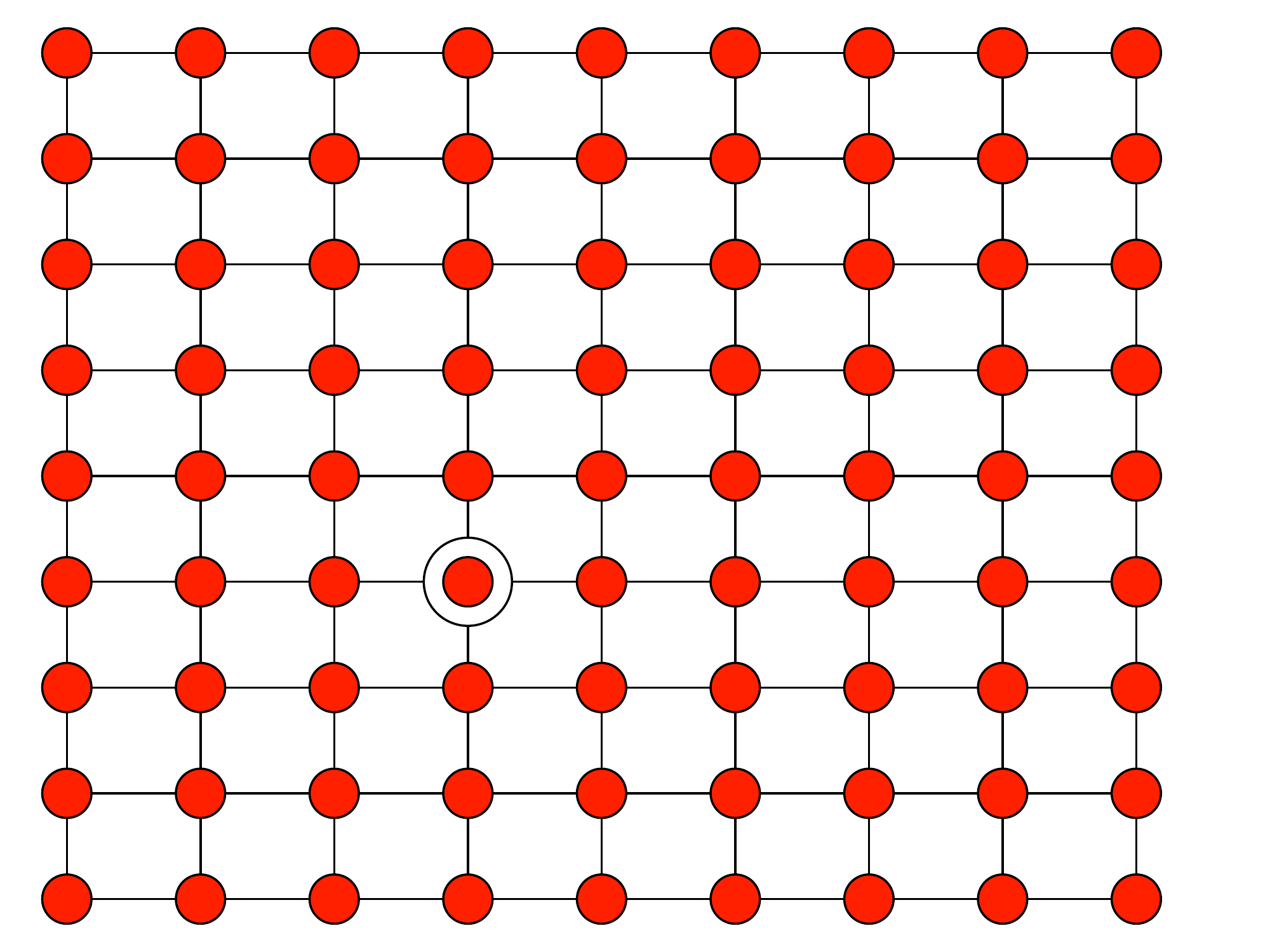}}}\hfill
\raisebox{-0.5\height}{\mbox{\includegraphics[width=0.48\columnwidth]{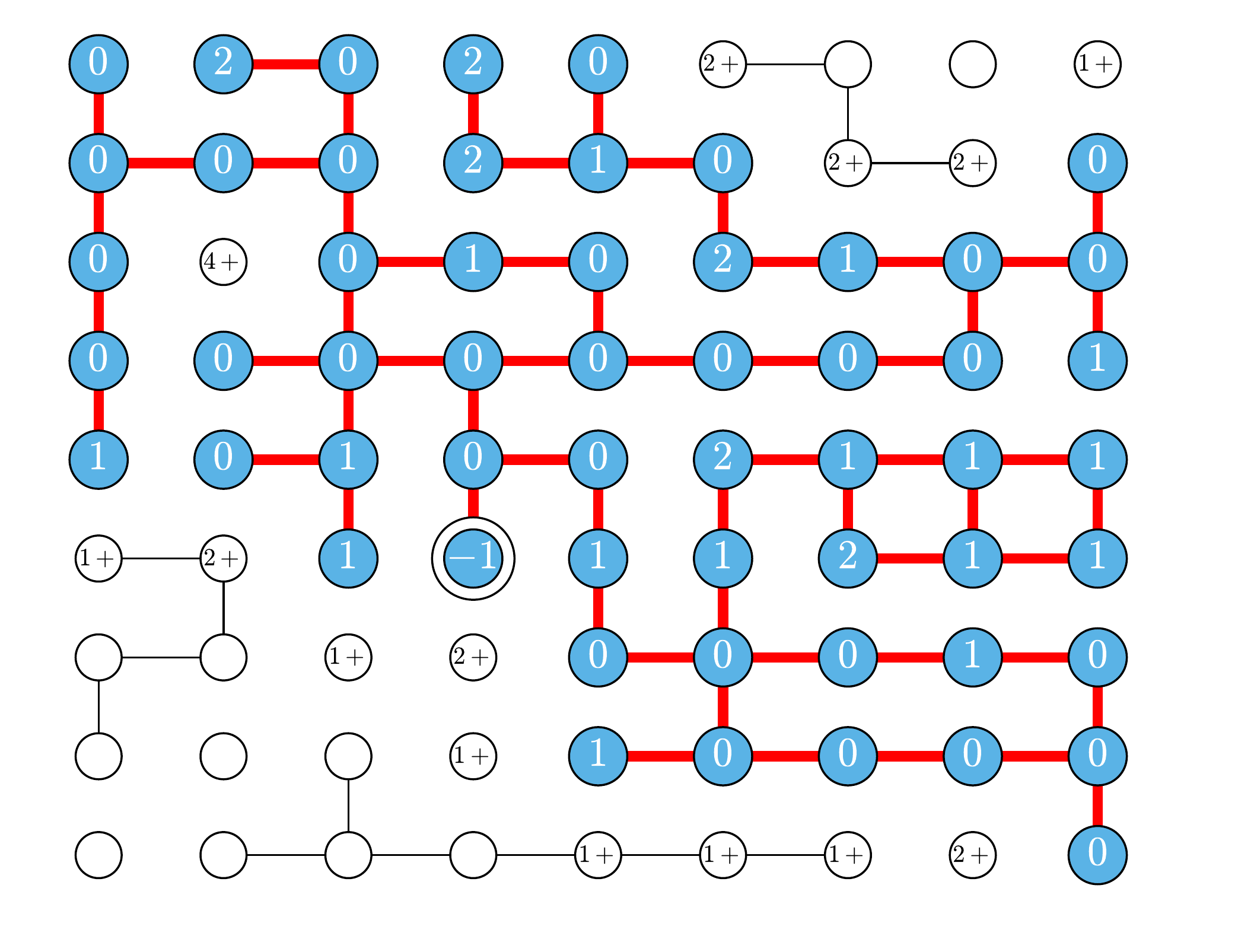}}}\hfill\mbox{}\\
\mbox{}\hfill\raisebox{-0.5\height}{\mbox{\includegraphics[width=0.48\columnwidth]{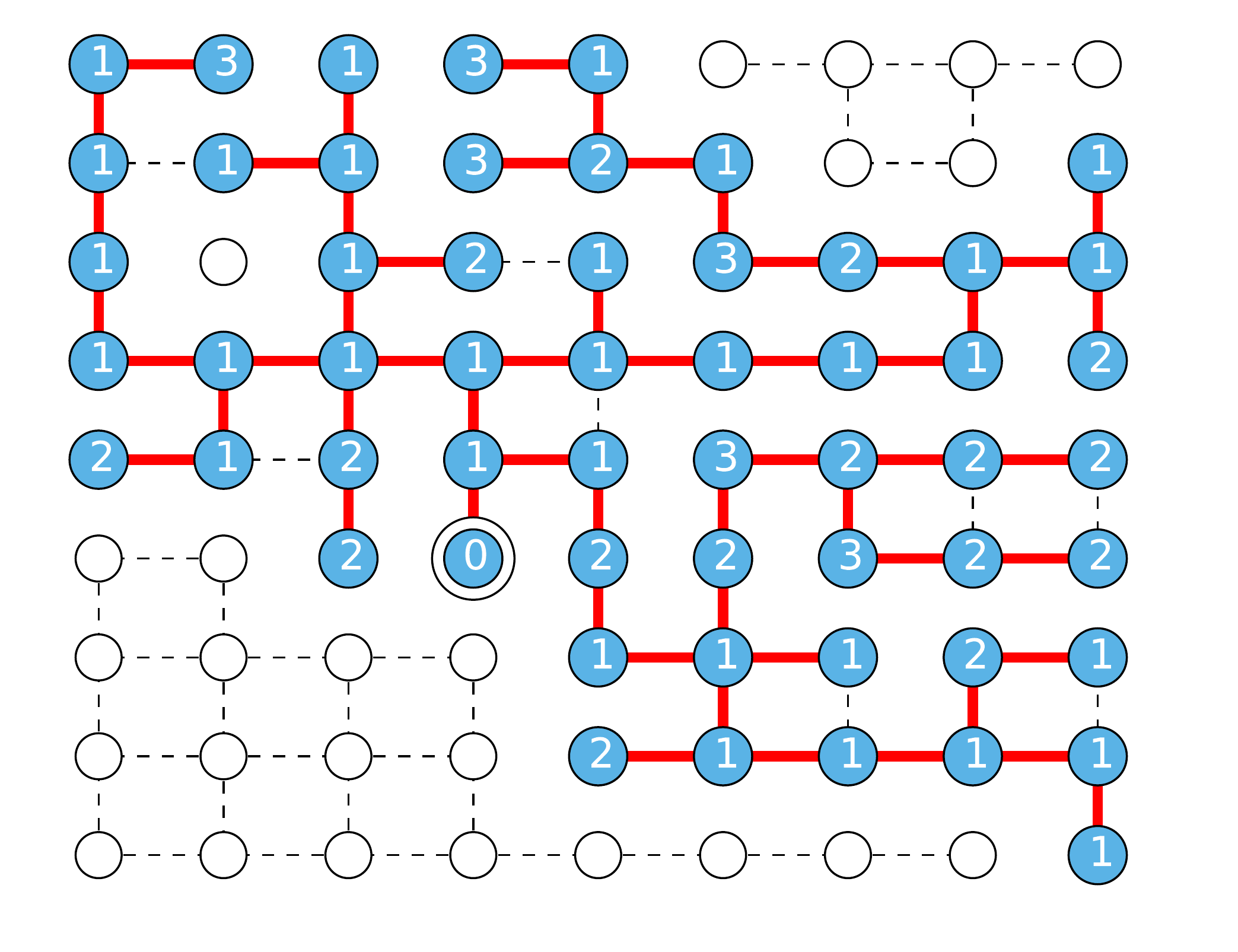}}}\hfill
\raisebox{-0.5\height}{\mbox{\includegraphics[width=0.48\columnwidth]{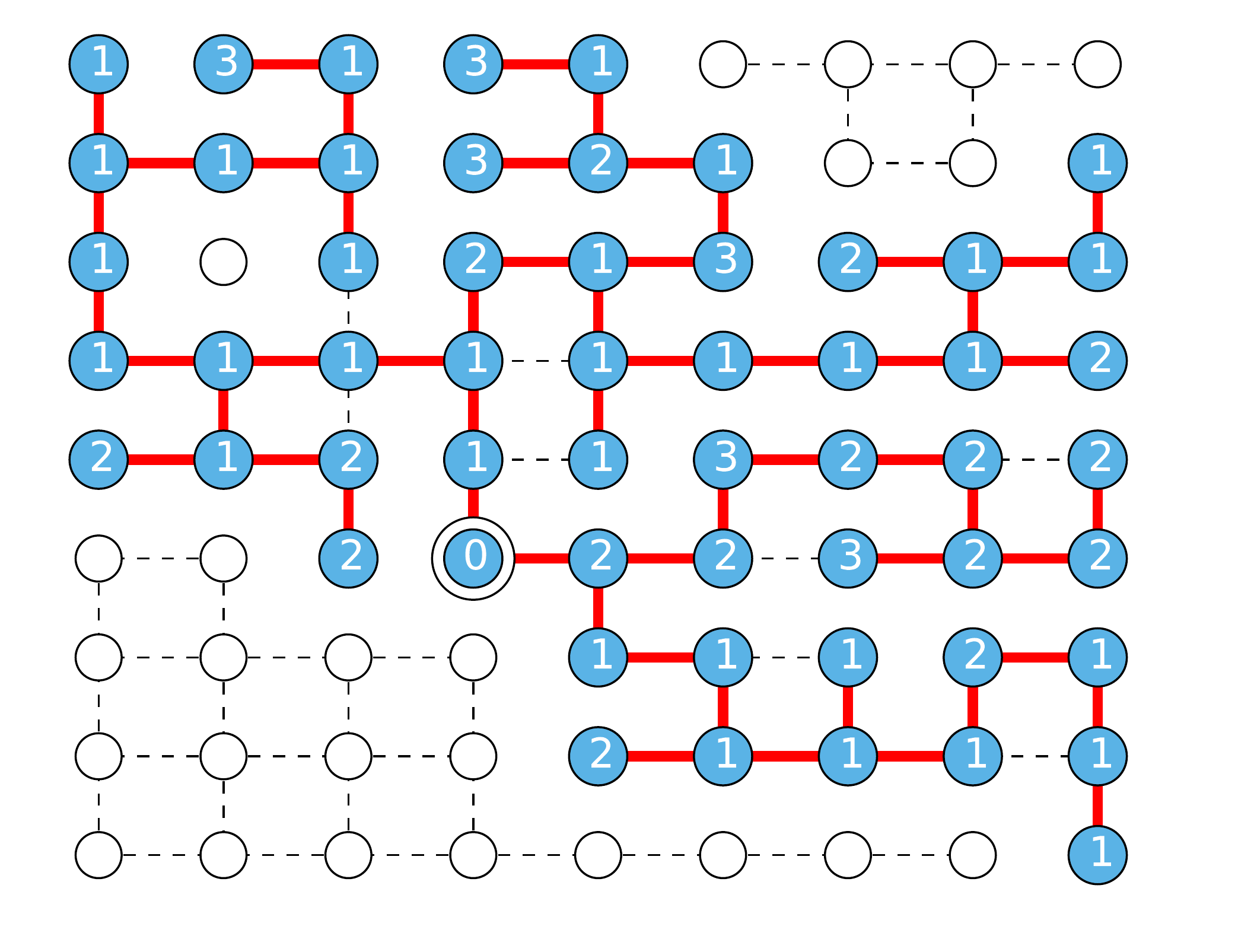}}}\hfill\mbox{}
\end{center}
\caption{(top left) Original network, with initial node highlighted.  (top right) Percolated network with component of initial node highlighted taking $p=0.51$.  Edges shown within a component are red, while edges in other components are black.  (bottom left) WTM outcome with thresholds found from the percolated network using a breadth-first-search for the WTM.  (bottom right) WTM outcome with thresholds from the percolated network using a depth-first search for the WTM.  In both WTM plots, the edges that were responsible for the activation of a node are shown in red.  Edges which were never considered are shown dashed in black.}
\label{fig:bp_vs_wtm}
\end{figure}

We will focus our attention just on identifying which nodes form connected components after bond percolation; we are not interested in which edges exist within the components.  In figure~\ref{fig:bp_vs_wtm} we compare a bond percolation approach to finding the component containing a particular node with a WTM approach for finding the same component.  We first perform bond percolation.  We then select an initial node (highlighted in the figure), and follow edges out from that node in the percolated network to find its component.  Nodes are labeled with $r$ where $r$ is the number of edges of the original network that were encountered (but deleted) \emph{prior to} an undeleted edge.  

We can think of this as being indistinguishable from selecting an initial node, following edges out from that node in some order, where each time an edge is considered, it is deleted with probability $1-p$ or followed with probability $p$.  The probability that the first $r$ edges to a node are deleted but the next is not is $p(1-p)^{r}$.

We compare this with the WTM with a threshold of $\tau_u = r_u+1$.  The activated nodes are identical to the component found using bond percolation.  In general, assigning nodes a threshold of $\tau$ where $\tau \geq 1$ is taken with probability $p(1-p)^{\tau-1}$ will yield a set of active nodes from an initially active node which come from the same distribution as the component of that node following bond percolation.  

In fact, we can generalize this approach to find all the components.   The steps in our process are to begin with a network, and assign thresholds using a geometric distribution:  for a threshold of $\tau$ the probability of $\tau$ is $p(1-p)^{\tau-1}$.  We then select a node and successively add nodes to its component once their threshold number of neighbors have been visited.  This process is likely to  terminate without exploring all nodes.  If this happens, we iteratively select a new node and add nodes to its component whenever their threshold number nodes have been visited (either in this stage or while building a previous component).  The resulting components match the components observed in bond percolation.  Nodes are activated exactly when they are added to a component in the bond percolation, and identifying in which iteration they are activated tells us which component they are part of.  The breadth-first search figure is the implementation of the WTM shown in Figure~\ref{fig:WTM_alg}, but we highlight that an alternate implementation with a depth-first search would yield the same outcomes.  As long as the \emph{initial} nodes of each pass are chosen in the same order, the precise details of the search algorithm do not determine which additional nodes belong to the component identified in that pass.

\begin{figure}[ht]
\includegraphics[width=0.5\columnwidth]{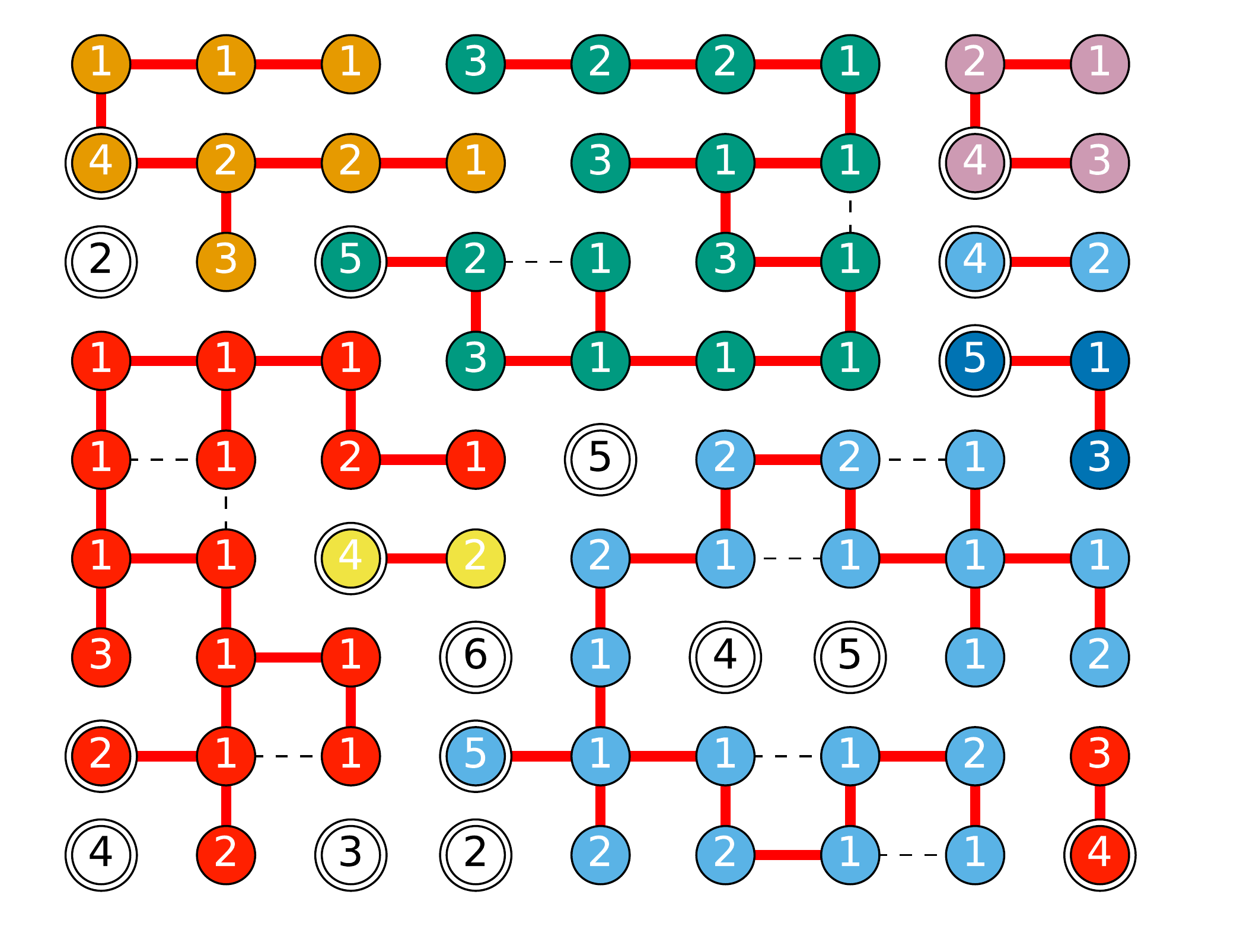}%
\includegraphics[width=0.5\columnwidth]{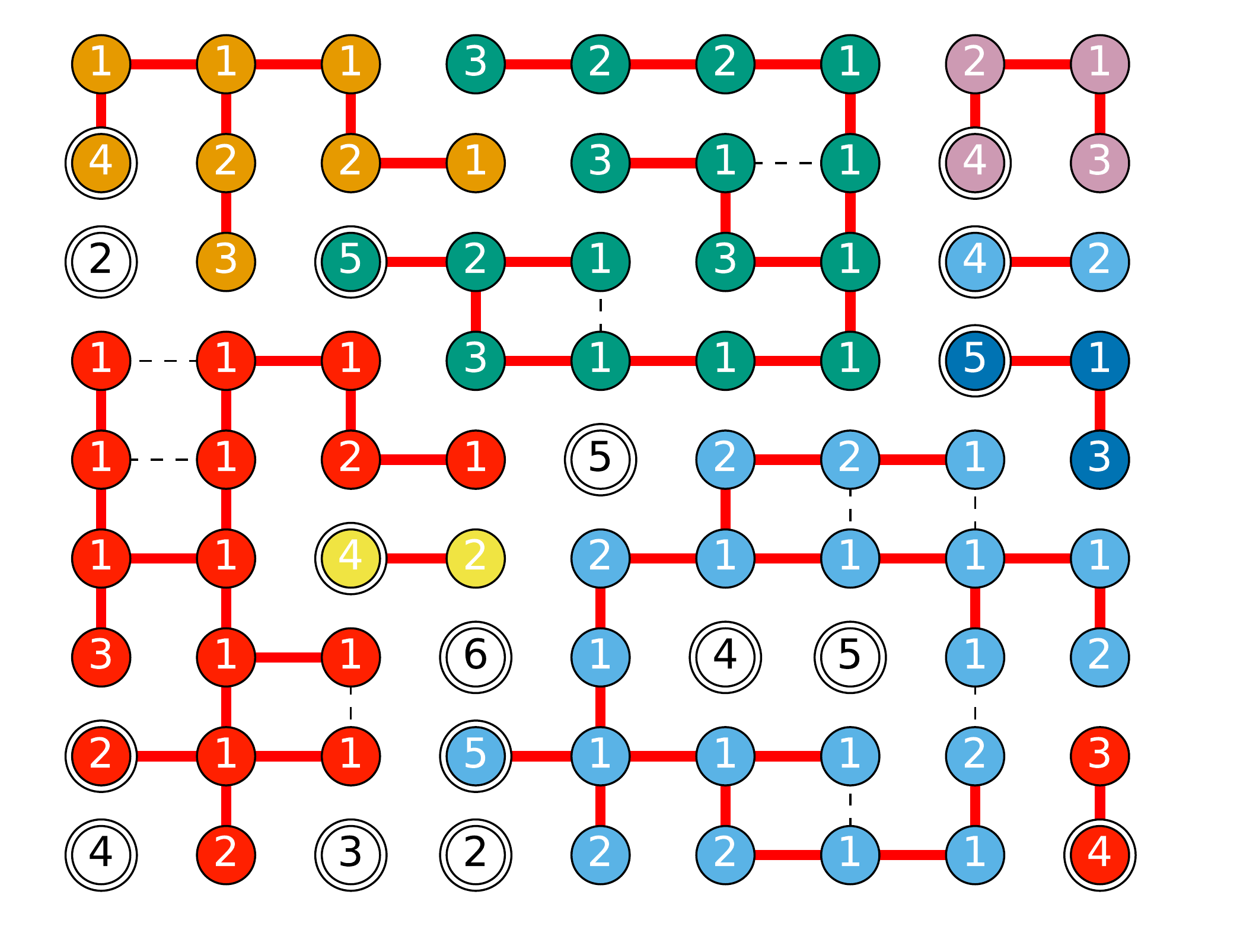}
\caption{Activated clusters found using depth-first (left) and breadth-first (right) searching using the WTM with a threshold of $\tau$ occuring with probability $p(1-p)^{\tau-1}$  (independently of $d$) and $p=0.51$ for a $9\times 9$ lattice.  The number at each node is its threshold.  The circled nodes are the initial nodes chosen for each cluster.  The bottom left node is chosen first, and its cluster traced out.  The next cluster is initialized by the bottom-most of the left-most remaining nodes.  Thick colored edges formed the final interaction that caused activation.  Non-existent edges failed to cause activation (but moved the node closer to its threshold).  Dashed black edges were not tested because both nodes were already active when the edge was considered.  The clusters remain the same for both search orders (but edges change).}
\label{fig:DfsVsBfs}
\end{figure}

To arrive at bond percolation, the thresholds for the WTM process are assigned from a geometric distribution.  It would be interesting to study whether a different distribution could be interpreted in the context of a generalized bond percolation.

\section{Discussion}
Many percolation processes have been studied in networks.  We have shown that site percolation, bootstrap percolation, $k$-core percolation, and the GEP are all special cases of the WTM.  In fact, the GEP we consider is equivalent to the WTM, and if we allow a node-specific threshold then both bootstrap and $k$-core percolation are also equivalent.  Which one should be considered the ``base'' model is a matter of personal choice.  

Bond percolation is closely related to the WTM, but to arrive at an equivalent model, the WTM assigns thresholds from a geometric distribution, activates a node, follows the WTM process to completion, and then activates another node.  The successive sets of activated nodes occur with the same probability as would be found in bond percolation.

We have further shown that generalizing the WTM to allow for a homogeneous transmission probability $T$ from active nodes to neighboring inactive nodes results in a model which can be thought of as a special case of the WTM.  Thus the potential space of models is not increased by this modification.

This commonality helps to explain why similar behaviors are observed and similar mathematical methods apply to these different processes.  

\section{Acknowledgments}
I thank Davide Cellai and James Gleeson for very useful conversations.

\appendix
\section{Appendix: Algorithm}
In figure~\ref{fig:WTM_alg} we give pseudocode for the WTM algorithm.  Other implementations are possible (this one is based on a breadth-first search, but for example, a depth-first search could also be used).  The choice of the function \textsf{dist\us{}func} which maps a randomly chosen weight from $(0,1)$ to a threshold allows us to match other percolation models.
\begin{figure}
\parbox{\columnwidth}{
\begin{algorithmic}
\Require Input network $G$, function generating numbers from a distribution \textsf{dist\us{}func},
and set of initially active nodes $I_0$.
\Ensure Set ActivatedNodes of activated nodes.
\begin{center}\rule{0.6\linewidth}{1pt}
\end{center}
\Function{WTM\us{}Assign\us{}Weights}{$G$}
\For{$u$ in $G$.nodes}
\State Assign weight[$u$] uniformly from $(0,1)$
\EndFor
\State \Return weight
\EndFunction
\begin{center}\rule{0.6\linewidth}{1pt}
\end{center}
\Function{WTM\us{}Assign\us{}Thresh}{$G$, \textsf{dist\us{}func}, weight, $I_0$}
\For{$u$ in $G$.nodes}
\If{$u$ in $I_0$}
\State thresh[$u$] $\gets 0$
\Else{}
\State thresh[$u$] $\gets$ \textsf{dist\us{}func}(G.degree($u$), weight[$u$])
\EndIf
\EndFor
\State \Return thresh
\EndFunction
\begin{center}\rule{0.6\linewidth}{1pt}
\end{center}
\Function{WTM\us{}Process}{$G$, thresh}
\State CurrentNodes $\gets$ set of nodes in $G$ with thresh $\leq 0$
\State ActivatedNodes $\gets$ set of nodes in CurrentNodes
\While{CurrentNodes is not empty}
\State NextNodes $\gets$ emptySet
\For{$u$ in CurrentNodes}
\For{$v$  in $G$.neighbors($u$)}
\If{$v \not \in $ ActiveNodes}
\State ActiveNodes.add($v$)
\State NextNodes.add($v$)
\EndIf
\EndFor
\EndFor
\State CurrentNodes $\gets$ NextNodes
\EndWhile
\State \Return ActiveNodes
\EndFunction
\begin{center}\rule{0.6\linewidth}{1pt}
\end{center}
\Function{WTM}{$G$, \textsf{dist\us{}func}, $I_0$}
\State weight = \textsf{WTM\us{}Assign\us{}Weights}($G$)
\State thresh = \textsf{WTM\us{}Assign\us{}Thresh}($G$,\textsf{dist\us{}func}, weight, $I_0$)
\State ActivatedNodes = \textsf{WTM\us{}Process}($G$, thresh)
\State \Return ActivatedNodes
\EndFunction
\begin{center}\rule{0.6\linewidth}{1pt}
\end{center}
\end{algorithmic}
}
\caption{The steps of the WTM algorithm.  The main algorithm is the final function given.  The other functions are called by the main algorithm.  First the weights are assigned randomly, and then the weights are mapped (deterministically) to a threshold and algorithm proceeds iteratively (and deterministically).  The appropriate choice of \textsf{dist\us{}func} allows us to select between the different models.}
\label{fig:WTM_alg}
\end{figure}

\bibliographystyle{plain}
\providecommand{\noopsort}[1]{}

\end{document}